\documentclass[twocolumn]{aastex63}
\pdfoutput=1 %for arXiv submission
\usepackage[caption=false]{subfig}

\begin{document}

\title{The Hubble PanCET program: Transit and Eclipse Spectroscopy of the Strongly Irradiated Giant Exoplanet WASP-76b}

\author{Guangwei Fu}
\author{Drake Deming}
\affiliation{Department of Astronomy, University of Maryland, College Park, MD 20742, USA; gfu@astro.umd.edu}

\author[0000-0003-3667-8633]{Joshua Lothringer}
\author{Nikolay Nikolov}
\author{David K. Sing}
\affiliation{Department of Physics and Astronomy, Johns Hopkins University, Baltimore, MD 21218, USA}

\author{Eliza M.-R. Kempton}
\author{Jegug Ih}
\affiliation{Department of Astronomy, University of Maryland, College Park, MD 20742, USA; gfu@astro.umd.edu}

\author[0000-0001-5442-1300]{Thomas M.\ Evans}
\affil{Kavli Institute for Astrophysics and Space Research, Massachusetts Institute of Technology, 77 Massachusetts Avenue, 37-241, Cambridge, MA 02139, USA}

\author{Kevin Stevenson}
\affiliation{Space Telescope Science Institute, 3700 San Martin Drive, Baltimore, MD 21218, USA}

\author[0000-0003-4328-3867]{H.R. Wakeford}
\affil{School of Physics, University of Bristol, HH Wills Physics Laboratory, Tyndall Avenue, Bristol BS8 1TL, UK}
 
\author{Joseph E. Rodriguez}
\author{Jason D. Eastman}
\affiliation{Center for Astrophysics | Harvard \& Smithsonian, 60 Garden Street, Cambridge, MA 02138, USA}

\author{Keivan Stassun}
\affiliation{Vanderbilt University, Department of Physics and Astronomy, 6301 Stevenson Center Lane, Nashville, TN 37235, USA}

\author{Gregory W. Henry}
\affiliation{Tennessee State University, Center of Excellence in Information Systems, Nashville, TN  37209, USA}

\author{Mercedes López-Morales}
\affiliation{Center for Astrophysics | Harvard \& Smithsonian, 60 Garden Street, Cambridge, MA 02138, USA}

\author{Monika Lendl}
\affiliation{Space Research Institute, Austrian Academy of Sciences, Schmiedlstr. 6, A-8042 Graz, Austria}

\author{Dennis M. Conti}
\affiliation{American Association of Variable Star Observers, 49 Bay State Road, Cambridge, MA 02138, USA}

\author{Chris Stockdale}
\affiliation{American Association of Variable Star Observers, 49 Bay State Road, Cambridge, MA 02138, USA}

\author{Karen Collins}
\affiliation{Center for Astrophysics | Harvard \& Smithsonian, 60 Garden Street, Cambridge, MA 02138, USA}

\author{John Kielkopf}
\affiliation{University of Louisville, 102 Natural Science Building, Louisville, KY 40292, USA}

\author{Joanna K. Barstow}
\affiliation{The Open University, Walton Hall, Kents Hill, Milton Keynes,  MK7 6AA, UK}

\author[0000-0002-1600-7835]{Jorge Sanz-Forcada}
\affil{Centro de Astrobiolog\'{i}a (CSIC-INTA),Spain}

\author{David Ehrenreich}
\author{Vincent Bourrier}
\author[0000-0002-2248-3838]{Leonardo A. dos Santos}
\affil{Observatoire astronomique de l'Universit\`e de Gen\`eve, 51 chemin des Maillettes 1290 Versoix, Switzerland}

\begin{abstract}

Ultra-hot Jupiters with equilibrium temperature greater than 2000K are uniquely interesting targets as they provide us crucial insights into how atmospheres behave under extreme conditions. This class of giant planets receives intense radiation from their host star and usually has strongly irradiated and highly inflated atmospheres. At such high temperature, cloud formation is expected to be suppressed and thermal dissociation of water vapor could occur. We observed the ultra-hot Jupiter WASP-76b with 7 transits and 5 eclipses using the Hubble Space Telescope (HST) and the Spitzer Space Telescope ($Spitzer$) for a comprehensive study of its atmospheric chemical and physical processes. We detected TiO and H$_2$O absorption in the optical and near-infrared transit spectrum. Additional absorption by a number of neutral and ionized heavy metals like Fe, Ni, Ti, and SiO help explain the short wavelength transit spectrum. The secondary eclipse spectrum shows muted water feature but a strong CO emission feature in Spitzer's 4.5\,$\mu$m band indicating an inverted temperature pressure profile. We analyzed both the transit and eclipse spectra with a combination of self-consistent PHOENIX models and atmospheric retrieval (ATMO). Both spectra were well fitted by the self-consistent PHOENIX forward atmosphere model in chemical and radiative equilibrium at solar metallicity, adding to the growing evidence that both TiO/VO and NUV heavy metals opacity are prominent NUV-optical opacity sources in the stratospheres of ultra-hot Jupiters. 

\end{abstract}
\nopagebreak

\section{Introduction}
Transiting exoplanets can offer us detailed insights into their atmospheres during the transit and eclipse phases. When transiting in front of the parent star, the limb of planetary atmosphere filters out a portion of the starlight. The amplitude of that effect varies with wavelength, depending on the composition of the atmosphere. The spectral features of the upper exoplanetary atmosphere ($\sim$1mbar) are thereby imprinted onto the stellar light. During the secondary eclipse, the planet passes behind the host star, and deep (10-100 mbar) thermal emission of the atmosphere can be measured via the total flux difference before and after the eclipse \citep{charbonneau05, deming05}. Both techniques have been used extensively in recent years to characterize exoplanetary atmospheric properties like chemical composition \citep{kreidberg15}, thermal structure \citep{stevenson17}, aerosols \citep{sing16} and hydrodynamical escape \citep{spake18, sing19}.

Most detectable exoplanetary spectral features produce only a few hundred ppm of signal over broad wavelength ranges \citep{deming13, fraine14, stevenson14a, wakeford17}.  High precision photometry is required to capture these small variations in the depth of transit and eclipse light curves. Indeed, since the first detection of sodium absorption in HD\,209458b made by \citep{charbonneau02} using the Hubble Space Telescope (HST), many atmospheric studies have used space telescopes, notably HST and Spitzer. Some recent ground based observations \citep{ehrenreich15, allart18, nikolov18, kirk20} have also successfully detected various atmospheric features such as water, sodium and helium. Chemical species that absorb in the very high atmosphere ($\sim 10$ scale heights) can cause a few thousand ppm excess transit depth within the narrow range of the absorption line profile core, which is often detectable from the ground despite additional noise from telluric contamination and changing weather conditions. 

Hot Jupiters are especially targets of interest for atmospheric characterization due to their inflated and highly irradiated atmospheres which produce strong detectable spectral features \citep{fortney08, mandell13}. Over a dozen hot Jupiters \citep{stevenson16, tsiaras18} have been studied in detail over the past decade and the results are highly intriguing yet complex \citep{fu17, sing16}. While some planets exhibit prominent water absorption features \citep{deming13, wakeford13}, others show significant aerosols presence in the upper atmosphere \citep{pont13}. Inverted temperature pressure profiles have also been observed \citep{haynes15, evans17} caused by optical absorbers such as TiO/VO \citep{fortney08, hubeny03}. In the ultra-hot ($>$2000K) Jupiters, even water can be disassociated and H- becomes an important opacity source \citep{arcangeli18, lothringer18, parmentier18, kitzmann18}. 

WASP-76b is a unique target with an equilibrium temperature of 2200K and a puffy atmosphere. Recent work has shown the existence of atomic sodium absorption \citep{seidel19, vonEssen20} and evidence for atomic iron condensing on the day-to-night terminator \citep{ehrenreich20}.  Here we present observations and modeling results that show heavy metals, $\mathrm{H}_2\mathrm{O}$ and TiO absorption in the transmission spectrum. The eclipse emission spectrum shows CO emission feature in the Spitzer's 4.5\,$\mu$m band with an inverted temperature pressure profile.

\section{Observations and Data Analysis}

We observed a total of 7 transits and 5 eclipses of WASP-76b with HST and Spitzer in multiple filters (Table \ref{observations}) ranging from 0.29 to 4.5\,$\mu$m. HST STIS/WFC3 and Spitzer IRAC all have unique detector systematics that require specialized data analysis pipelines \citep{deming13, nikolov13, deming15, wakeford16}. Fortunately, as the main instruments used to characterize exoplanetary atmospheres in the past decade, robust custom data analysis methods have been developed to extract near photon-limited noise spectra \citep{zhou17}.

\begin{table*}[t]
\centering
\begin{tabular}{cccccc}
\multicolumn{6}{c}{\textbf{\large{WASP-76b transit observations}}}\\
\hline\hline 
&	Grism/Filter	&	Visit 1	&	Visit 2	   &   GO Program ID & PI \\   
\hline 
HST STIS	&	G430L	&	2016-11-16	&	  2017-01-17	&	14767	& L\'{o}pez-Morales $\&$	Sing	\\
HST STIS	&	G750L	&	2017-02-19	&		&	14767	&	L\'{o}pez-Morales $\&$ Sing	\\
HST WFC3	&	G141	&	2015-11-26	&		&	14260	&	Deming	\\
Spitzer	&	IRAC 3.6	&	2017-05-04	&	 2018-04-22	&	13038	&	Stevenson	\\
Spitzer	&	IRAC 4.5	&	2017-04-16	&		&	13038	&	Stevenson	\\ 
\hline 
\smallskip \\
\multicolumn{6}{c}{\textbf{\large{WASP-76b eclipse observations}}}\\
\hline\hline
&	Grism/Filter	&	Visit 1	&	Visit 2   &   GO Program ID & PI \\   
\hline 
HST WFC3	&	G141	&	2016-11-03	&		&	14767	&	 L\'{o}pez-Morales $\&$	Sing	\\
Spitzer	&	IRAC 3.6	&	2016-03-22	&		&	12085	&	Deming	\\
Spitzer	&	IRAC 3.6	&	2017-05-04	&	2018-04-22	&	13038	&	Stevenson	\\
Spitzer	&	IRAC 4.5	&	2016-04-01	&		&	12085	&	Deming	\\
\hline 

\end{tabular}
\caption{A list of our 7 transit and 5 eclipse observations of WASP-76b.}
\label{observations}
\end{table*}

\subsection{Companion Star \& EXOFASTv2 Fit}

\begin{figure*}
    \centering
    \includegraphics[width=1\textwidth]{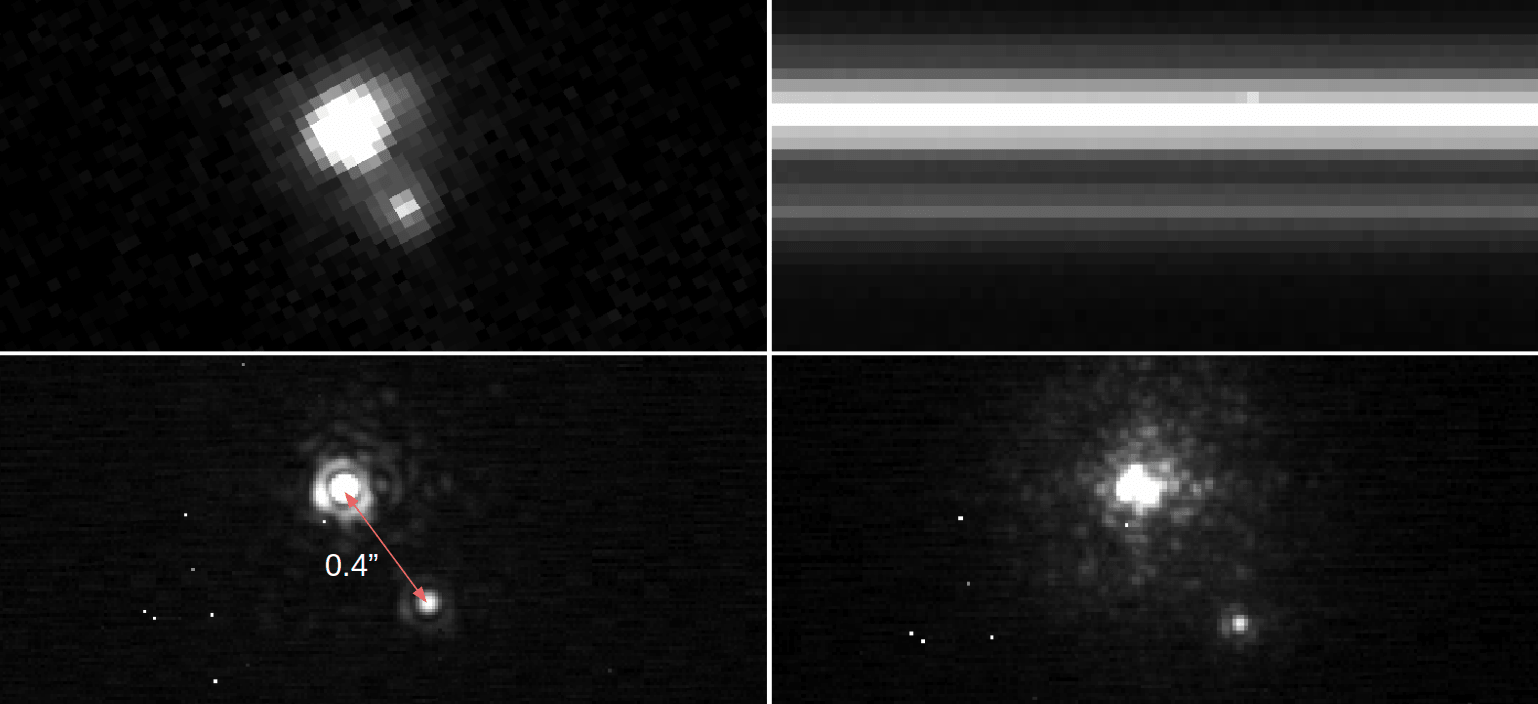}
    \caption{Resolved images of spectrum of the WASP-76 binary system, obtained with HST STIS F28X50LP (Top Left), G750L (Top Right), Keck-AO NIRC2 Brackett-gamma (Bottom Left) and J-Cont (Bottom Right).}%
    \label{companion}
\end{figure*}

WASP-76A has a companion star. WASP-76B was first discovered by \citep{wollert18} through lucky imaging with a separation of 0.425" $\pm$ 0.012" and position angle of 216.9$^\circ$ $\pm$ 2.93$^\circ$. Due to the small separation, light from WASP-76B is well mixed with WASP-76A in our HST spatial scan spectrum which causes a dilution effect on the extracted planet spectrum \citep{crossfield12}. To correct for this dilution effect, the companion stellar spectral type needs to be determined, and the extra flux contribution removed. The temperature of WASP-76A is 6250 $\pm$ 100K \citep{west16} and the updated distance from GAIA \citep{gaia18} is 195.31 $\pm$ 6.03\,parsecs. There are a total of three spatially resolved images of the WASP-76 system in the archive, taken with different filters (Table~\ref{observations}) using the Space Telescope Imaging Spectrograph (P.I.s: David Sing \& Mercedes L\'{o}pez-Morales), and Keck-AO with NIRC2 (P.I.: Brad Hansen), all shown in Figure~\ref{companion}. To determinate the spectral type of WASP-76B, we performed a two-component SED fit \citep{rodriguez19} for WASP-76 and we determine the radius and temperature of WASP-76B to be $R_{\star} = 0.795 \pm{0.055} R_{\Sun}$ and $T_{eff} = 4850 \pm{150K}$ which we then used as the prior in a EXOFASTv2 \citep{eastman19} global analysis (Table \ref{exofast}). Within the EXOFASTv2 fit, the host star parameters are constrained using the we use the MESA Isochrones and Stellar Tracks (MIST) stellar evolution models \citep{paxton15, dotter16, choi16}. For the EXOFASTv2 fit, we included six new light curves (Fig. \ref{fig:exofast_transits}) from EulerCAM \citep{ehrenreich20}, Hazelwood and MVRC observations in addition to the transit and RV data used in the discovery paper \citep{west16} to refine and update the system ephemeris.

With the best-fit radius and effective temperature for both WASP76A and WASP76B, we can then use PHEONIX stellar models to calculate the flux contribution from both stars and the dilution effect of transit and eclipse depth can be corrected as follows:
\[Corrected\ depth = Measured\ depth * (1+\frac{F_B}{F_A})\]
where $F_B$ and $F_A$ are the flux contribution from the companion and the primary star at a given wavelength range. Since the companion star is spatially resolved at different levels with being mostly resolved in STIS spectra while completely blended in at Spitzer bands, we purposefully choose larger aperture sizes at all wavelength when extracting stellar spectra to ensure all companion flux contributions are included. Finally, the dilution is applied across the entire transit and eclipse spectra for consistent correction.

To propagate the uncertainties on the effective temperature of both stars into dilution factors and the final planet spectrum, we adopted the bootstrapping method used in \citep{stevenson14b} by generating 10000 PHOENIX stellar models for each star with $T_{eff}$ randomly sampled from a Gaussian distribution based on the $T_{eff}$ uncertainty. By calculating the corresponding dilution factors for each PHOENIX model pair, we obtain a 10000 sample size distribution of dilution factors at each wavelength bin. The final dilution factors are the median values of each distribution and the uncertainties will be the corresponding one sigma values which are then propagated into the reduced planet transit and eclipse spectra.   

\begin{figure}
    \centering
    \includegraphics[width=0.5\textwidth]{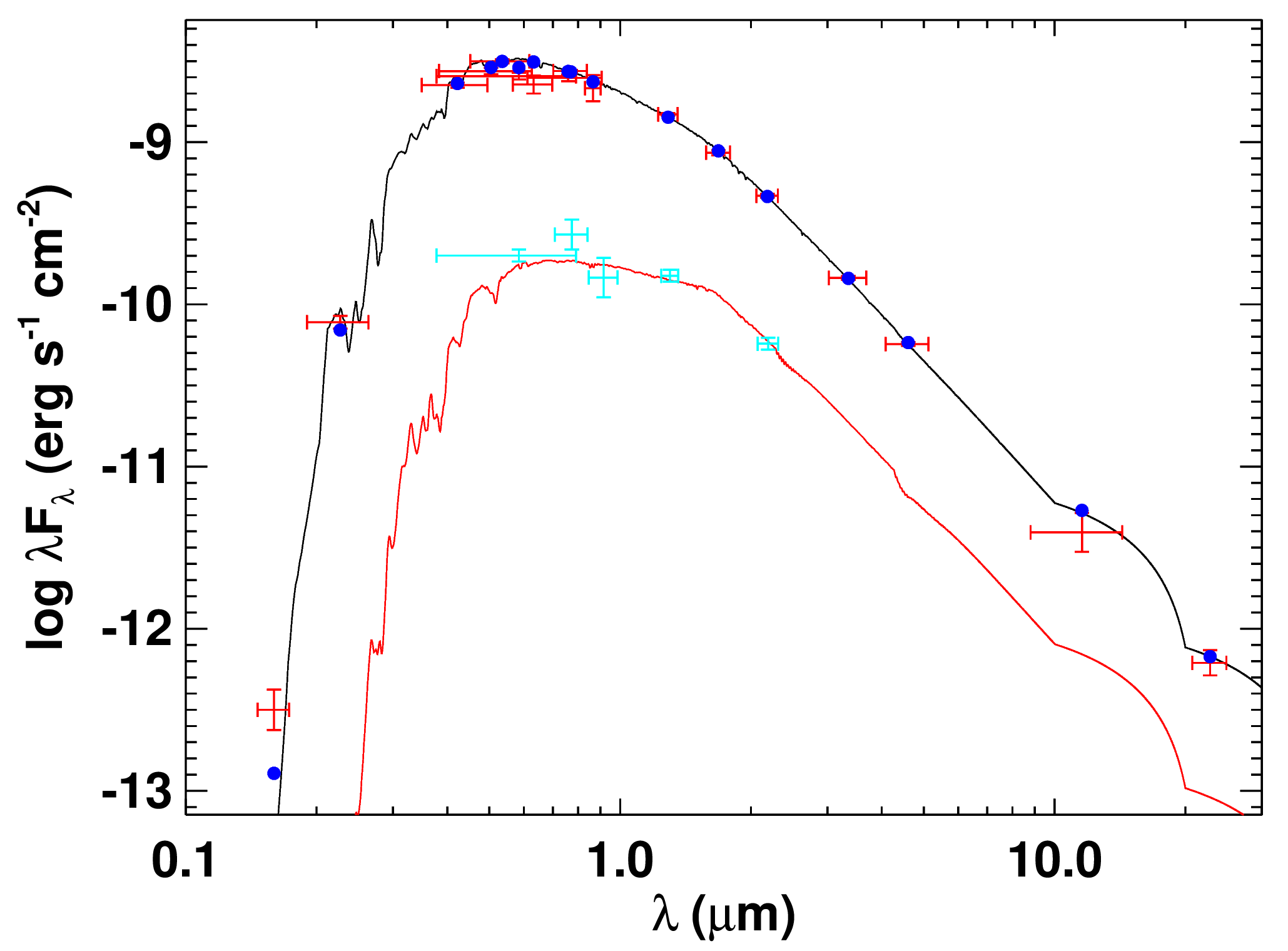}
    \caption{Two-component SED fit for WASP-76A (black) and WASP-76B (red) with the blue points as integrated fluxes and cyan points as spatially resolved flux measurements. }%
    \label{sed}
\end{figure}

\begin{table*}[t]
\centering
\begin{tabular}{c|ccccc}
\hline\hline 
& STIS F28X50LP	&	i band	&	z band	&	J-Cont	&	Br gamma \\   
\hline 
Wavelength range ($\mu m$)	&	0.54 - 1	&	0.662 - 0.836	&	0.777 - 1.097	&	1.203 - 1.223	&	2.024 - 2.292	\\
$\Delta$ mag	&	2.57	&	2.51	&	2.85	&	2.49	&	2.28	\\
$\Delta$ mag error	&	0.02	&	0.25	&	0.33	&	0.01	&	0.01	\\

\hline 
\end{tabular}
\caption{Measured flux ratio between WASP-76A and WASP-76B in 5 different bands from HST STIS, \cite{wollert18} and KECK-AO NIRC2.}
\label{table1}
\end{table*}

\subsection{HST STIS G430L \& G750L}

We observed WASP-76b in transit with 2 visits using HST/STIS G430L and 1 visit using the G750L grating (Table \ref{observations}). Both gratings were observed using the ACCUM mode with the 50X2 aperture to minimize any slit losses. CCD subarray of 128 x 1024 pixels was used to reduce readout time and maximize observing efficiency. Each frame has an exposure time of $\sim$148 seconds and each orbit has $\sim$16 exposures. The combination of these two gratings provided a complete wavelength coverage from 2900 to 10300 \AA. One prominent source of systematics in STIS light curves comes from the orbital motion of the telescope during the observations \citep{nikolov14}. As the telescope orbits between the day side and night side of the Earth, it experiences thermal expansion and contraction. This effect manifests as a varying observed flux as a function of telescope orbital phase.

\begin{figure*}
\centering
  \includegraphics[width=\textwidth,keepaspectratio]{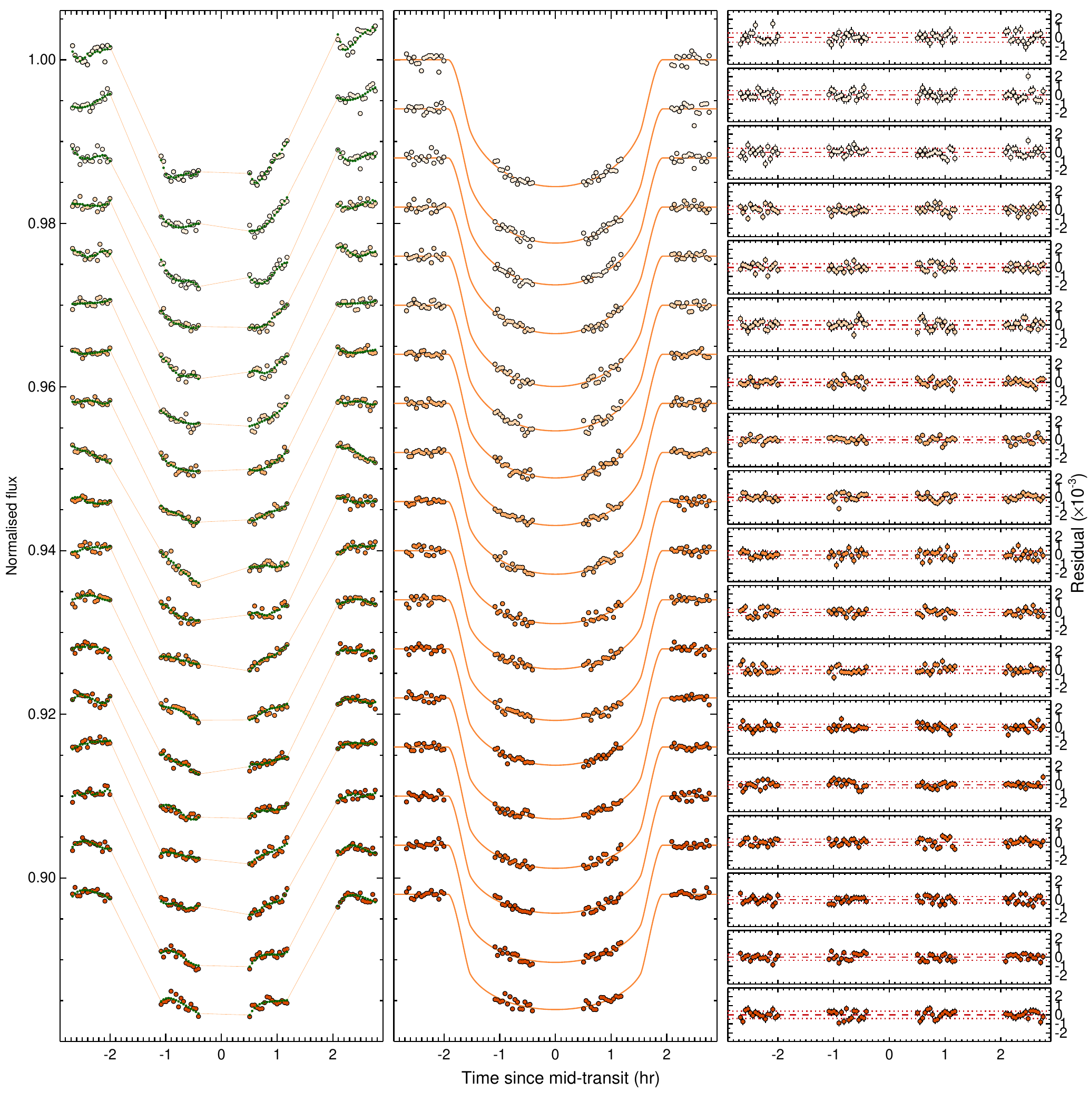}
  \caption{HST STIS G430L visit one light curves for each spectral channel. Left panel: raw light curves with evident systematics as a function of telescope orbital phase. Middle panel: De-trended light curves overplotted by the best-fitting transit models. Right panel: corresponding residual for each spectral channel with the dotted lines showing the 1$\sigma$ standard deviation.}
  \label{fig:V1_G430L}
\end{figure*}

\begin{figure*}
\centering
  \includegraphics[width=\textwidth,keepaspectratio]{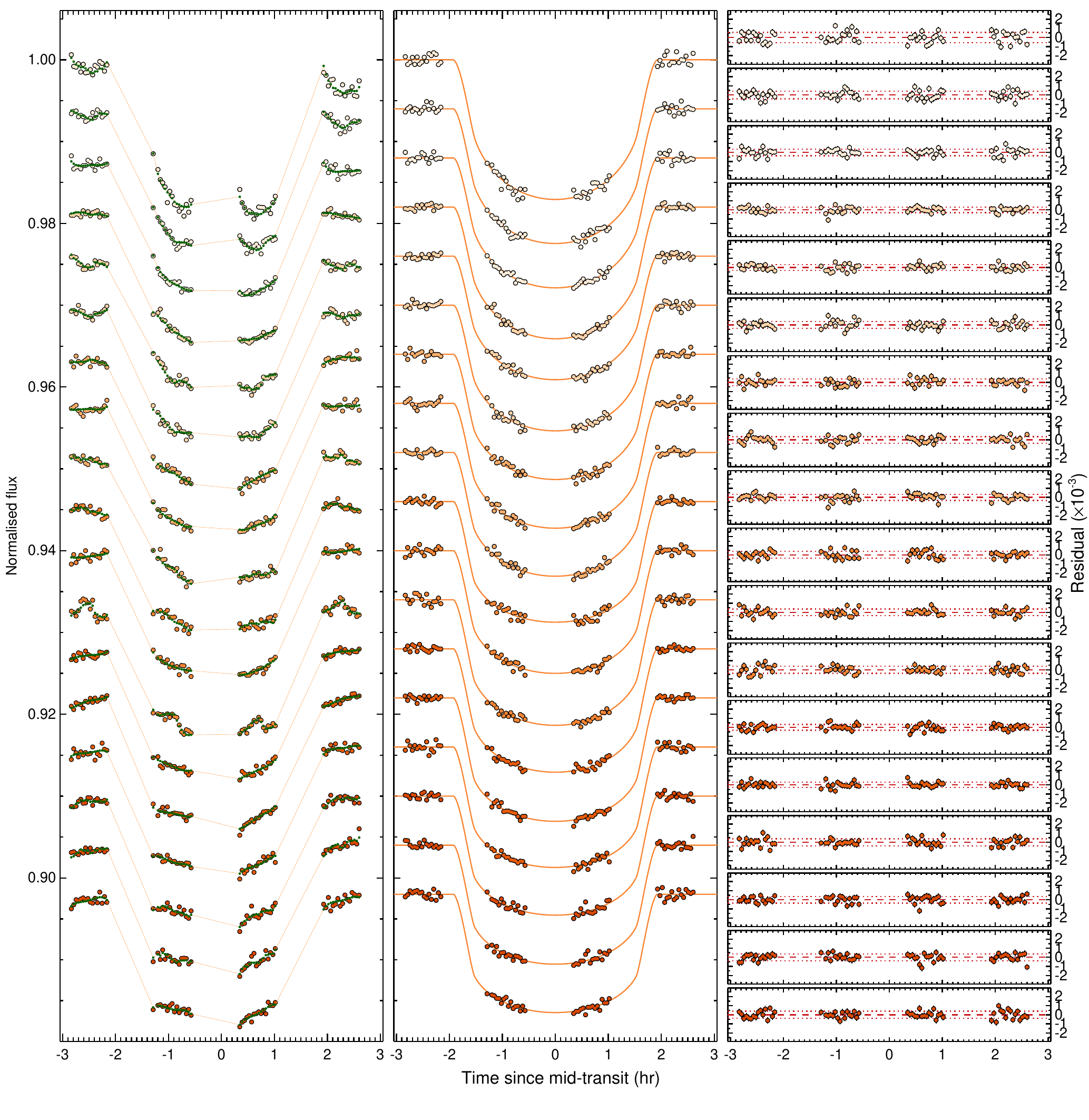}
  \caption{Same as Fig \ref{fig:V1_G430L} but for HST STIS G430L visit 2.}
  \label{fig:V2_G430L}
\end{figure*}

\begin{figure*}
\centering
  \includegraphics[width=\textwidth,keepaspectratio]{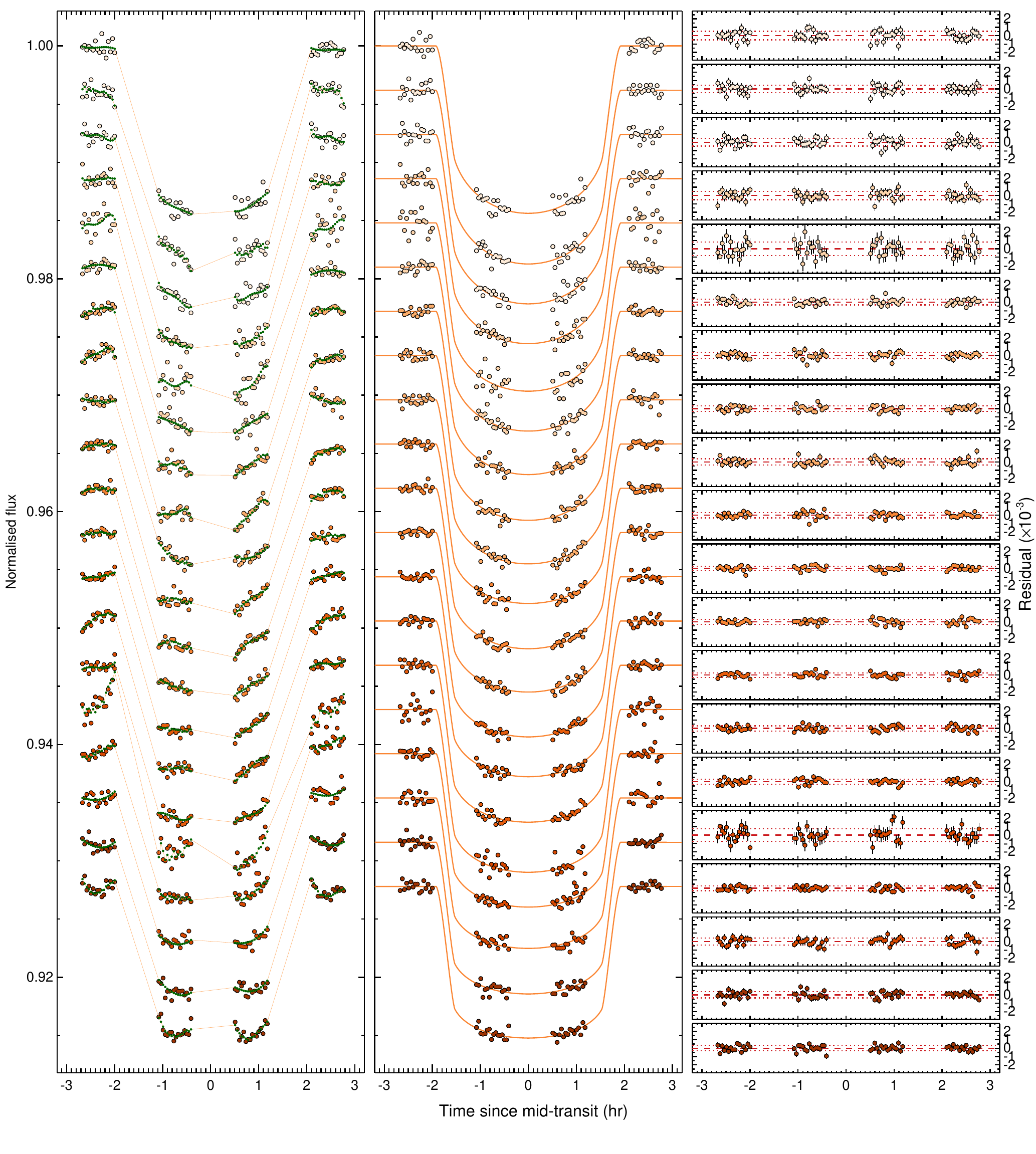}
  \caption{Same as Fig \ref{fig:V1_G430L} but for HST STIS G750L.}
  \label{fig:G750L}
\end{figure*}

Our data analysis process follows the standard methodology detailed in \cite{sing11} and \cite{nikolov15}. We fit the STIS transit light curves using a combination of transit and instrument systematics models. The transit model is based on the analytic formula developed by \cite{mandel02}, and the systematics model is a fourth-order polynomial of the telescope orbital phase, a linear time term and wavelength shift ($\omega$) for each frame. Orbital inclination and $a/R_{star}$ are both fixed at the best-fit values derived in this paper during the fit. For limb darkening we calculate the relevant coefficients with ATLAS stellar models in the same way as detailed in \cite{nikolov15}. The raw, corrected light curves and corresponding residuals for all three visits are shown in Figures \ref{fig:V1_G430L}, \ref{fig:V2_G430L} and \ref{fig:G750L}.

\subsection{HST WFC3 G141}

We observed both transits and eclipses using HST/WFC3 G141 in spatial scan mode to maximize photon-collecting efficiency \citep{deming13}.  All frames used SPARS10 and NSAMP=16, with an exposure time of $\sim$104 seconds, and a forward and backward scan to maximize observing efficiency. Due to occultation of the telescope by the Earth, a $\sim$45 min gap exists between every HST orbit. In total, there are 5 orbits per visit and $\sim$19 spectra per orbit. Two orbits are pre-transit, two are in-transit, and one is post-transit. 

The automatic CalWF3 pipeline does not include spatial scan mode, therefore additional processing is required before extracting the 1D spectra. We followed the standard procedures of background subtraction and energetic particle removal by flagging outliers relative to the median value along the vertical scan direction \citep{wakeford13}. Next, we corrected for the wavelength shift of each spectrum in the horizontal direction. To calculate the sub-pixel level shifts between each frame, we first summed each frame in the vertical direction to obtain a 1D spectrum and normalized it by its own median flux. Then we used $scipy.interpolate.interp1d$ function to interpolate normalized flux of each 1D spectrum in the wavelength direction relative to its pixel positions. Next we applied sub-pixel shifts to each 1D spectrum relative to a reference spectrum and calculated the shifts by minimizing the normalized flux differences between them. Finally we applied the calculated shifts on every 1D spectrum to obtain the wavelength shifts corrected 1D spectra. The hydrogen Paschen-beta line at 1.28\,$\mu$m in the star is used to establish the zero-point of the wavelength calibration. 

HST/WFC3 time series spectra often exhibit a ramp-like systematic shape when observing bright stars in high-cadence \citep{wilkins}. This effect is attributed to charge trapping in the WFC3 HgCdTe infrared detector \citep{kreidberg14N, zhou17}. As initial photons arrive at the beginning of each orbit, some charge carriers can be trapped by impurities in the detector and cause lower readout signals. When all available traps are filled during the orbit, the measured signals asymptotically approach a constant level (Fig \ref{fig:transit_wl}). The double-ramp shape per orbit is due to differences in exposure timing and telescope pointing between forward and backward scan. The timings for when each pixel receives light are different in forward and backward scan, and that can affect the ramp shape.  Moreover, the illumination pattern on the pixel grid is slightly different from forward to backward scan.  Since each pixel has a different number of charge traps, a constant offset in measured flux can occur when different portions of the detector are illuminated by forward and backward scan.  

\subsubsection{Satellite contamination}

\begin{figure}%

\subfloat[Frame ID: icy002wpq]{%
  \includegraphics[clip,width=\columnwidth]{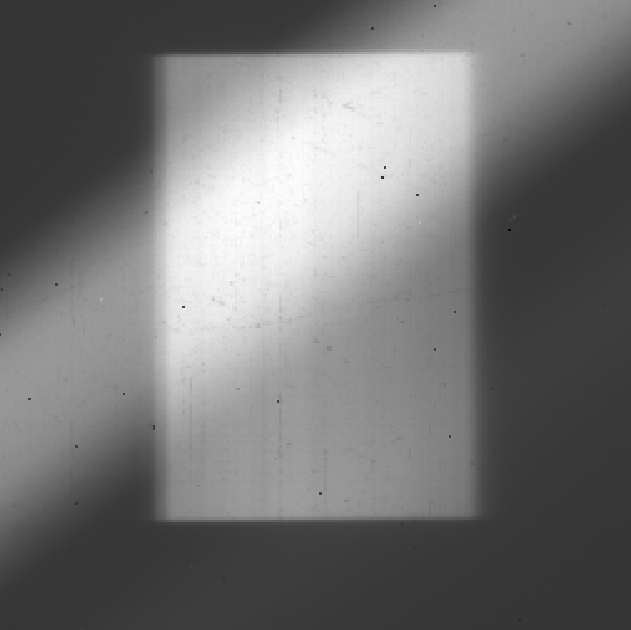}%
}

\subfloat[Frame ID: icy002xfq]{%
  \includegraphics[clip,width=\columnwidth]{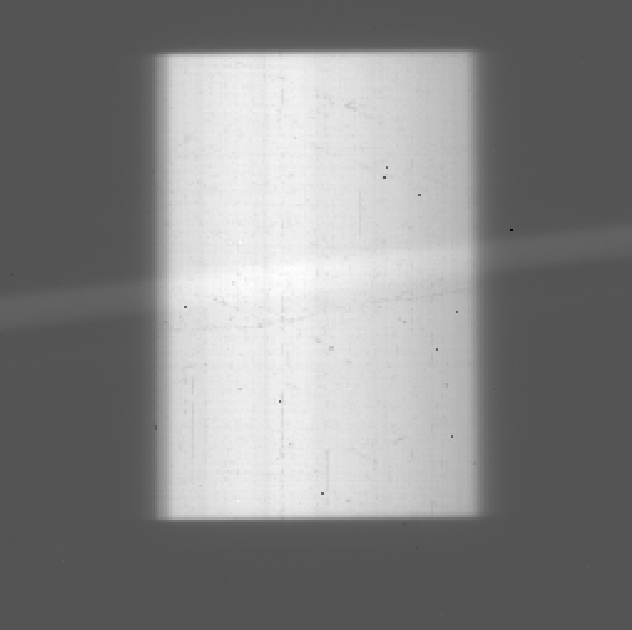}%
}

\caption{Satellite crossing contamination frames}%
\label{satellite}%
\end{figure}

During the analysis of the transit data, we discovered two frames (Fig \ref{satellite}) that were contaminated by defocused Earth-satellite crossing events. The first satellite crosses the frame diagonally (see Fig \ref{satellite}(a)) leaving a broad bright strip which contaminates the spectrum in a wavelength-dependent fashion. The extra photons from the satellite significantly distort the ramp shape of the third orbit's white light transit curve (see Fig \ref{fig:transit_wl}), because they rapidly populate large number of charge traps. This causes the decay-down as opposed to the ramp-up shape as extra persisting signals were measured in all subsequent frames of the orbit. The diagonal crossing of the satellite results in more contamination on the shorter wavelength end of the spectrum than the longer wavelength end. Consequently, the white light transit curve cannot be used as a template to correct for all wavelength channels. We decide to discard all remaining frames in the third orbit after the first satellite crossing. The second satellite crossing is much fainter and had negligible effect on the subsequent spectra in the fourth orbit, so we only discard the frame with the second satellite itself.

\subsubsection{Ramp correction using RECTE}

After removing satellite-contaminated exposures, we use the Ramp Effect Charge Trapping Eliminator (RECTE) algorithm developed by \cite{zhou17} to mitigate the ramp effect. RECTE is a physically motivated model based on detector charge trapping properties. For more detailed description of RECTE see \citep{zhou17} and the online documentation\footnote{https://recte.readthedocs.io }. One major advantage of RECTE compared to other ramp-effect correcting methods based on template and fitting of empirical functions is the capability to correct for the first orbit of the observations. The first orbit has often been discarded in past analyses due to its extreme ramp shape comparing to the subsequent four orbits \citep{deming13, kreidberg15}. Recovering that additional out-of-transit orbit allows us to better determine the baseline flux and obtain more precise transit depth values. 

\begin{figure}
  \includegraphics[width=0.5\textwidth,keepaspectratio]{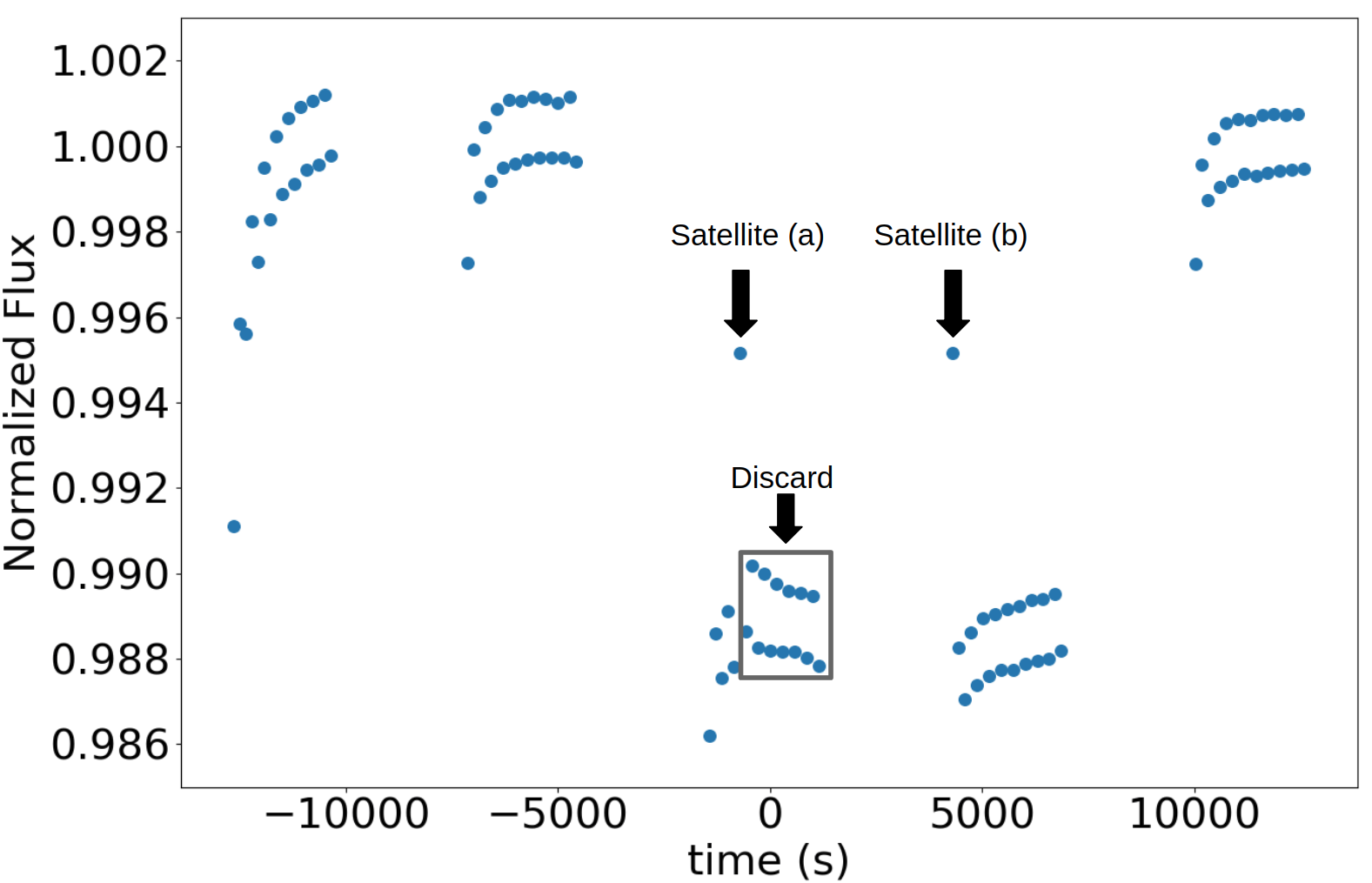}
  \caption{The effect of satellite crossing contamination on the white light transit curve. The two satellite contaminated data points have very high flux and were set to fixed values to show the timing of the events. Satellite crossing (a) was significantly more severe than satellite crossing (b) and distorted the ramp shape for the third orbit. We decided to discard the frames after satellite crossing (a), see text. The upper and lower sets of points are due to spectra vertical shifts during the forward versus backward scan.}
  \label{fig:transit_wl}
\end{figure}

We used the BATMAN light curve model \citep{kreidberg15b} in combination with RECTE to measure the transit depth at each spectral bin. Orbital inclination and $a/R_{star}$ were both fixed at the best-fit values derived in this paper during the fit. We calculated the relevant limb darkening coefficients with ATLAS stellar models the same way as the STIS dataset. There are five free parameters from RECTE: intrinsic flux ($f$), slow ($E_{s,tot}$), and fast ($E_{f,tot}$) charge traps populations, slow ($\eta_s$) and fast ($\eta_f$) charge trapping efficiency. Together they model the varying exponential ramp effect from the charge trapping process in the HST/WFC3 detectors. The slight vertical shift from forward and backward scans cause an observed flux difference between adjacent (Fig \ref{fig:transit_wl}) exposures which is corrected through fitting a constant offset value. There is also a linear visit-long slope which is fit with two slope coefficients for forward and backward scans. Given our re-fit of the orbital parameters that determine the shape of the transit, the BATMAN fit has two free parameters, the transit center time and transit depth. Therefore, a total of 10 free parameters were used in the MCMC to fit for the white light transit. The transit center time from the white light fit is adopted when subsequently fitting transit curves at each wavelength.

\begin{figure}
  \includegraphics[width=0.5\textwidth,keepaspectratio]{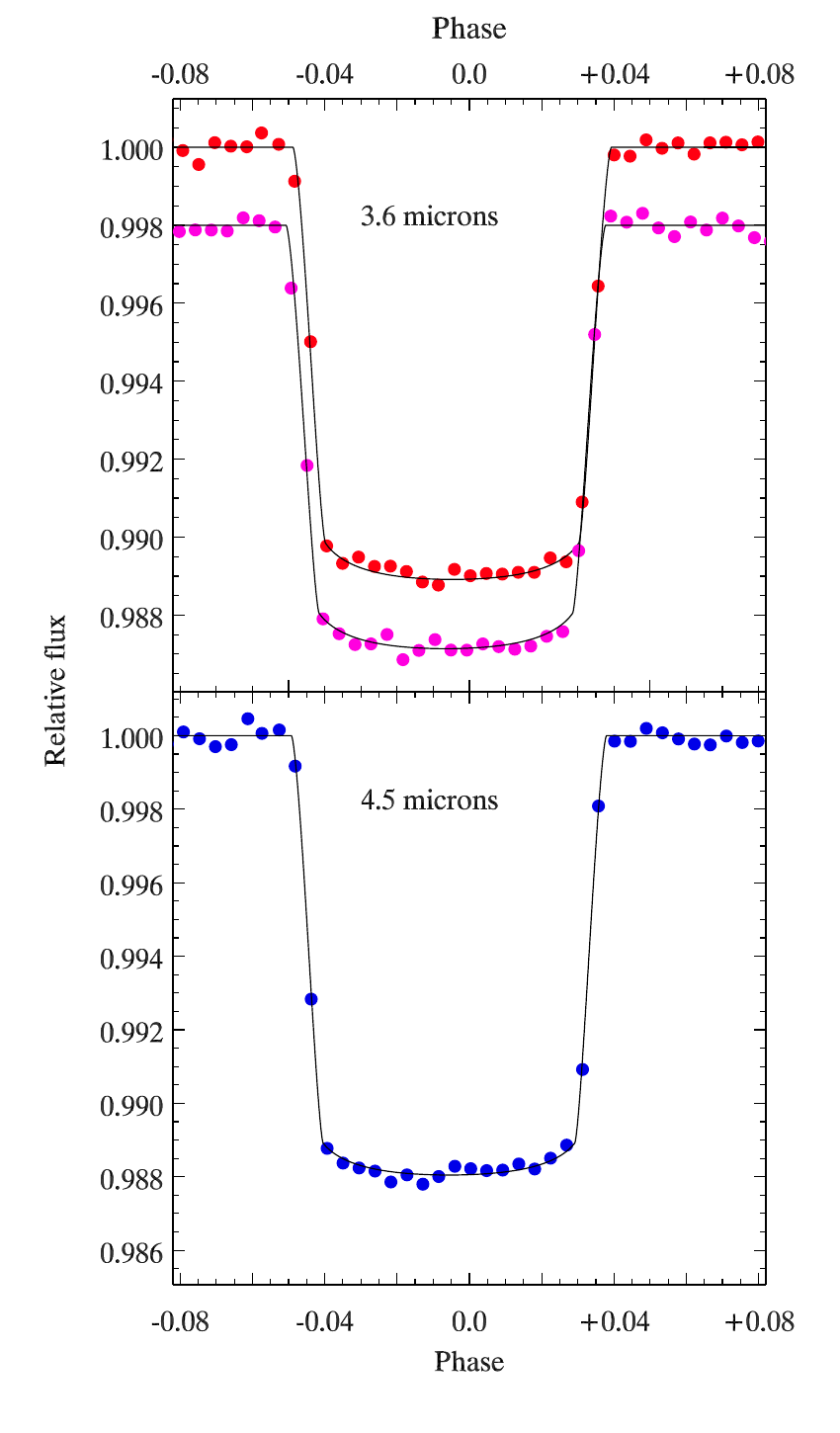}
  \caption{Spitzer transit light curves in 3.6 and 4.5 $\mu m$ after systematic correction.}
  \label{fig:W76_spitzer_transits}
\end{figure}

\begin{figure*}
\centering
  \includegraphics[width=\textwidth,keepaspectratio]{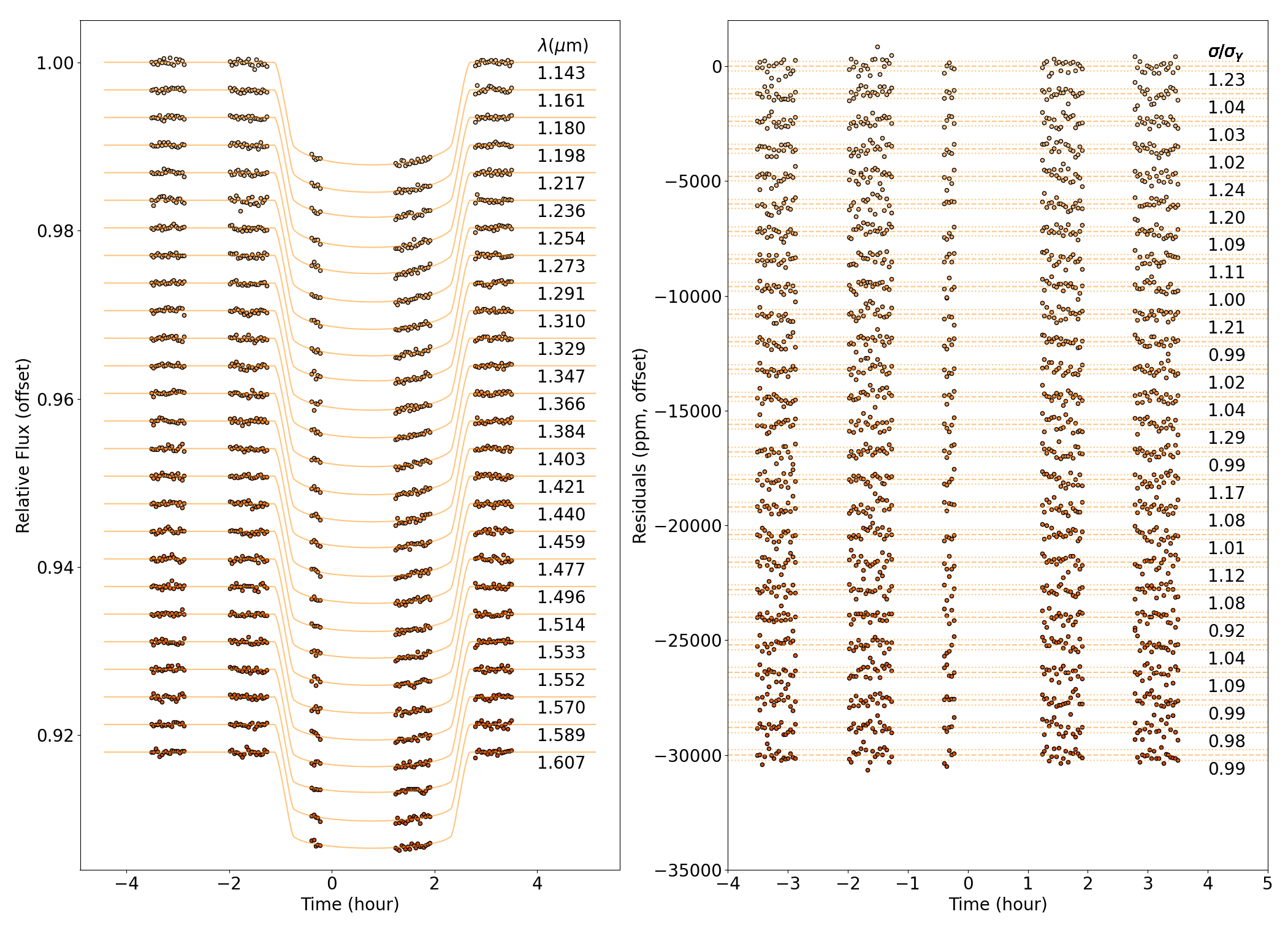}
  \caption{ HST WFC3 G141 spectral bin transit lightcurves after ramp-effect correction using RECTE (left) and corresponding residuals (right). The dotted lines in the residual plot represent expected photon noise.}
  \label{fig:transit_all_wvl}
\end{figure*}

\begin{figure*}
\centering
  \includegraphics[width=\textwidth,keepaspectratio]{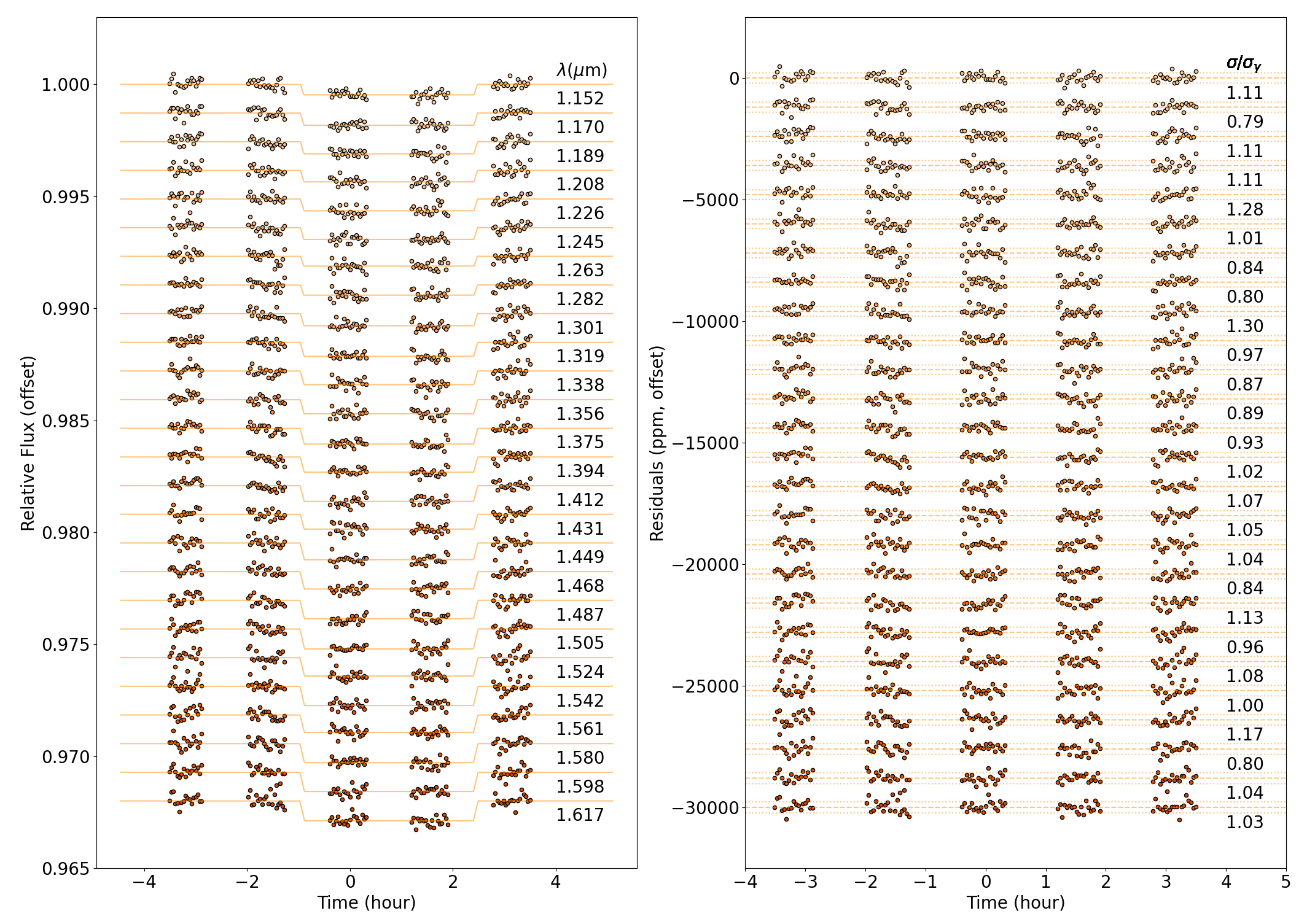}
  \caption{HST WFC3 G141 spectral bin eclipse light curves after ramp-effect correction using RECTE (left) and corresponding residuals (right). The dotted lines in the residual plot represent expected photon noise.}
  \label{fig:eclipse_all_wvl}
\end{figure*}

\subsection{Spitzer IRAC}

We observed transits and eclipses of WASP-76b with Spitzer at 3.6 and 4.5\,$\mu$m (Table \ref{observations}). Unlike HST, Spitzer is able to continuously observe targets for the entire transit and eclipse duration. One eclipse at each of 3.6 and 4.5\,$\mu$m are reported by \citet{garhart20}, and we do not re-analyze those data here. We here analyze the transit data, and two additional eclipses at 3.6\,$\mu$m from program 13038. Our analysis of two additional eclipses at 3.6\,$\mu$m followed the exact same procedures used by \citet{garhart20}, in fact the same codes (implemented by D.D.).  The new eclipse depths are included in Table~\ref{Spitzer_table}. 

Spitzer's primary systematic effect comes from intra-pixel sensitivity variations coupled to pointing jitter, overlaid by temporal ramps.  We correct for this combination of systematic effects using the pixel-level decorrelation (PLD) technique developed by \cite{deming15}, with the implementation of the fit being the same as described by \citet{garhart20}. PLD takes advantage of the total flux conservation within the aperture containing the star, and utilizes the relative flux contribution of individual pixels as basis vectors in the fit. This technique eliminates the need for finding the centroid position of the star while being capable of effectively removing red noise and flat-fielding inaccuracies. 

Our solutions for the Spitzer transit depths incorporate quadratic limb darkening coefficients calculated for the Spitzer bands by \citet{claret13}. These produce excellent agreement with the observed transit curves.  Given that limb darkening is a minimal effect at Spitzer's wavelengths, we adopt the Claret coefficients without further perturbation. Our initial procedure was to also freeze the orbital parameters at previously-determined values, since our experience with other data shows that this simple method usually produces excellent agreement with the shape of Spitzer's observed transit curves.  However, atmospheric characterization can be sensitive to alternate treatments of the orbital parameters \citep{alexoudi18}. Given also that we find some differences between the transit depths observed at 3.6- versus 4.5\,$\mu$m, and between two transits at 3.6\,$\mu$m, we explored other treatments of the orbital parameters. We used independent Gaussian priors for the two parameters that most affect the transit shape (orbital inclination and $a/R_s$), based on the discovery results from \citet{west16}. Those fits produced transit depths that differed minimally from fits that froze the orbital parameters at the \citet{west16} values. Those differences (orbital priors minus orbital freeze), were 157 and 34 ppm in $R_p^2/R_s^2$ for the two transits at 3.6\,$\mu$m, and -93 ppm at 4.5\,$\mu$m. Our best-fit values of inclination and $a/R_s$ differed from \citet{west16} by less than $1\sigma$. We also explored freezing the orbital parameters at the values derived in this paper, noting that our values for inclination and $a/R_s$ are within $1\sigma$ of \citet{west16}. Those transit depths differed from our initial values by 84 and 3 ppm at 3.6\,$\mu$m, and 128 ppm at 4.5\,$\mu$m.

In the various solutions for Spitzer transit depths described above, differences persist between 3.6 and 4.5\,$\mu$m, and between the two transits observed at 3.6\,$\mu$m.  Those differences are minimized by our default solutions, i.e. freezing the orbital parameters at the values given by \citet{west16} and solving for $R_p^2/R_s^2$.  Given that the orbital parameters we derive in this paper are closely consistent with \citet{west16}, we adopt our default solutions for transit depths. Those values are listed in Table~\ref{Spitzer_table}, and the best-fit transit times are included. Figure~\ref{fig:W76_spitzer_transits} illustrates the fits, after removal of the systematic effects, and binning the data for visual clarity.

\begin{table*}[]
\centering
\caption{Transit and eclipse times and depths for WASP-76b in the Spitzer bands. These are "as observed" transit/eclipse depths, not corrected for dilution by the companion star. The two eclipses from GO 12085 (PI: Deming) were published in \citep{garhart20} and therefore not included here.}
\begin{tabular}{llll}
%\newline
Wavelength  &  Event & BJD(TDB)  & Depth (ppm)  \\
\hline \hline
3.6\,$\mu$m & Transit & 2457877.915709$\pm$0.000163  &  10496$\pm$66    \\
3.6\,$\mu$m & Transit & 2458230.840367$\pm$0.000145  &  10315$\pm$49    \\
4.5\,$\mu$m & Transit & 2457859.815112$\pm$0.000181  &  11399$\pm$82    \\
3.6\,$\mu$m & Eclipse & 2457877.01558$\pm0.00067$    &   2883$\pm$96    \\
3.6\,$\mu$m & Eclipse & 2458229.93999$\pm0.00056$    &   3086$\pm$88    \\
\end{tabular}
\label{Spitzer_table}
\end{table*}

\section{PHOTOMETRIC OBSERVATIONS OF WASP-76}

We acquired a total of 208 out-of-transit observations of WASP-76 during five
recent observing seasons, not including several transit observations each year,
with the Tennessee State University (TSU) Celestron 14-inch (C14) automated 
imaging telescope (AIT) at Fairborn Observatory 
\citep[see, e.g.,][]{h1999,ehf2003}.  The AIT uses an STL-1001E CCD camera 
from Santa Barbara Instrument Group (SBIG); all exposures were made 
through a Cousins R filter.  Each observation consisted of 3--10 
consecutive exposures on WASP-76 and several comparison stars in the same 
field of view.  The individual frames were co-added and reduced to 
differential magnitudes – i.e. WASP-76 minus the mean brightness of 
seven constant comparison stars. Further details of our observing, reduction, 
and analysis techniques can be found in \citet{sing+2015}.

The photometric observations are summarized in Table \ref{ground}.  Column~4 lists the 
yearly standard deviations of the observations from their seasonal means; these
values are consistent with the precision of a single observation, as 
determined from the comparison stars.  Our SBIG STL-1001E CCD camera 
suffered a gradual degradation during the 2017-18 observing season, resulting 
in the loss of data from that season.  The camera was replaced with another 
SBIG STL-1001E CCD to minimize instrumental shifts in the data.  Nonetheless, 
there appears to be a shift in the seasonal-mean differential magnitudes, given
in column~5, of several milli-magnitudes between the third and fourth observing
seasons.  Otherwise, the night-to-night and year-to-year variability in 
columns 4 \& 5 show that WASP-76 is constant on both time scales to the limit 
of our precision. 

\begin{figure}
\includegraphics[width=0.5\textwidth,keepaspectratio]{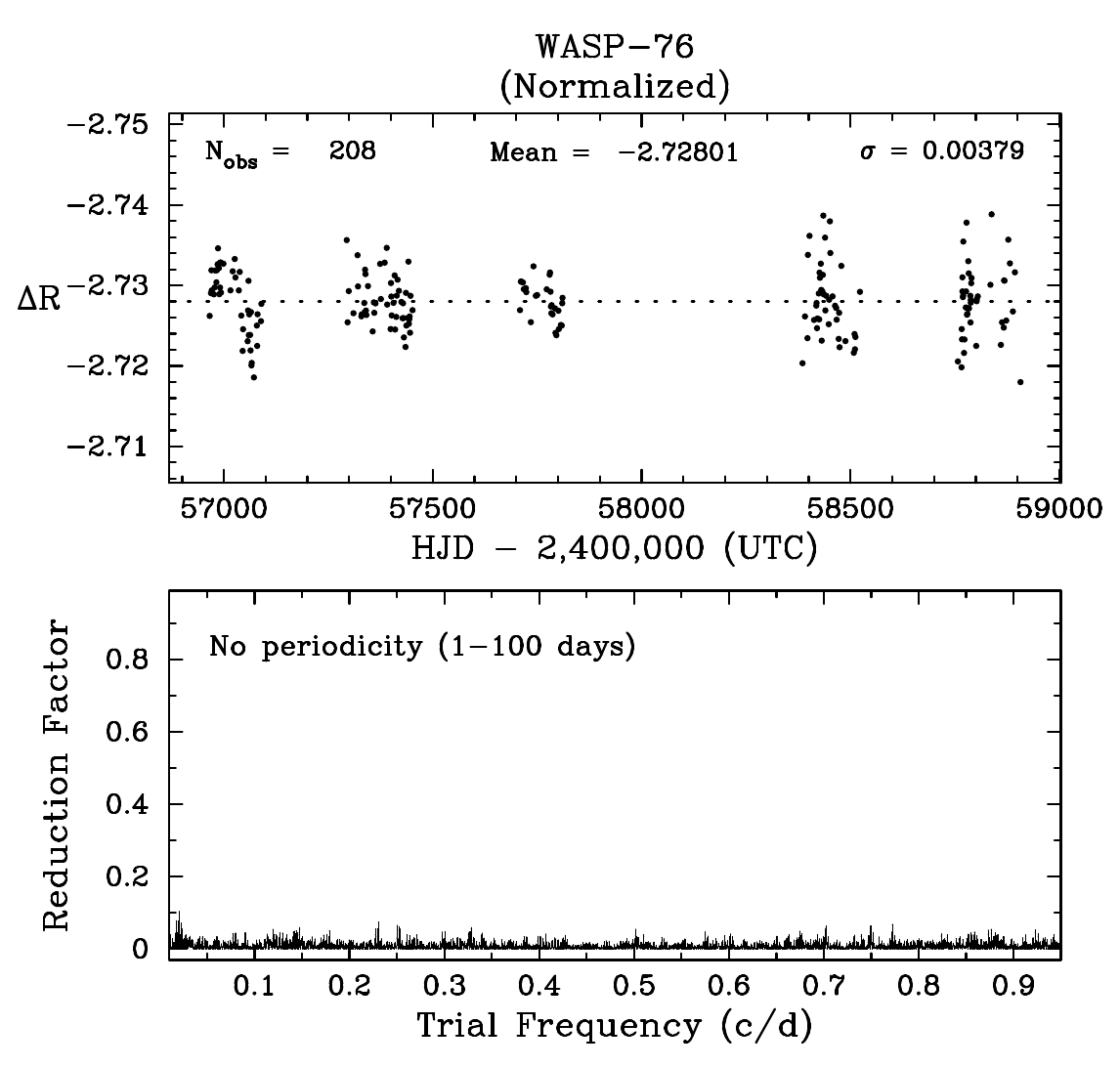}
    \caption{$Top$:  AIT photometry of WASP-76 between 2014 and 2020 but lacking
    the 2017-18 observing season. The observations have been normalized so that 
    all observing seasons have the same mean as the first season.  
    $Bottom$:  Frequency spectrum of the normalized observations showing the lack 
    of any significant periodicity between 1 and 100 days.}
    \label{fig:ground_monitor}
\end{figure}

The complete WASP-76 data set is plotted in the top panel of Figure \ref{fig:ground_monitor}, where the data have been normalized so that each seasonal-mean differential magnitude is the same as the first observing season. This removes any year-to-year variability in the comparison stars as well as long-term variability in 
WASP-76, if any. The bottom panel shows the frequency spectrum of our complete data set (note the absence of the 2017-18 observing season) and gives no evidence for any coherent periodicity between 1 and 100 days, as expected from the lack of variability shown in Table \ref{ground}. 

Our data were observed during a three years period. If the star is variable, we will suffer constant offsets in transit and eclipse depth between data taken at different times. The long term photometric monitoring of WASP-76 with no detection of any periodicity on short timescales allows us to confirm features in the planet spectra are not caused by any short term stellar variability. However, this does not rule out longer term variability causing potential offsets between observations separated by longer than a year since we have normalized each seasonal mean flux level to the first season. 

\begin{deluxetable*}{ccccc}
%\tabletypesize{\small}
%\tablewidth{0pt}
\tablecaption{SUMMARY OF AIT PHOTOMETRIC OBSERVATIONS OF WASP-76 \label{ground}}
\tablehead{
\colhead{Observing} & \colhead{} & \colhead{Date Range} &
\colhead{Sigma} & \colhead{Seasonal Mean} \\
\colhead{Season} & \colhead{$N_{obs}$} & \colhead{(HJD $-$ 2,400,000)} &
\colhead{(mag)} & \colhead{(mag)} \\
\colhead{(1)} & \colhead{(2)} & \colhead{(3)} &
\colhead{(4)} & \colhead{(5)}
}
\startdata
 2014-15  &  44 & 56965--57089 & 0.0040 & $-$2.7280  \\
 2015-16  &  51 & 57293--57451 & 0.0030 & $-$2.7301  \\
 2016-17  &  28 & 57708--57810 & 0.0024 & $-$2.7267  \\
 2018-19  &  42 & 58384--58522 & 0.0045 & $-$2.7346  \\
 2019-20  &  43 & 58756--58906 & 0.0045 & $-$2.7355  \\
\enddata
\end{deluxetable*}

\section{Comparison with previous studies}
We have compared our reduced transit spectrum with previous studies \citep{vonEssen20, edwards20} and we believe the major discrepancies come from the different approaches used for the satellite contaminated frames and the companion star dilution correction. For the WFC3 transit dataset we removed all frames in the second orbit after the satellite crossing and \cite{edwards20} only removed the satellite crossing frames themselves. Imperfect correction for the lingering extra flux (Fig \ref{fig:transit_wl}) induced by the satellite will result in a smaller transit depth. So we decided to adopted a more conservative approach to discard those frames. 

For the dilution correction, \cite{vonEssen20} fitted two Gaussian functions to the STIS 2D spectral images at each wavelength and then subtract the companion flux contribution. \cite{edwards20} used the WFC3 simulator Wayne to model the companion star flux contribution based on the reported K band delta magnitude and stellar parameters from \cite{bohn20}. 

Our approach is different as discussed in section 2.1, we fit for the companion star SED based on the observed photometric data points and uniformly apply the resulting dilution factors to STIS, WFC3 and Spitzer spectra. Our approach avoids the need to customizing for instrument specific systematics when correcting for the companion flux contribution. We are also able to propagate the uncertainties from the companion star stellar parameters into the final transmission and emission spectra of the planet consistently across all wavelength. As a result, our errorbars on the final spectra are larger than reported in previous studies \citep{vonEssen20, edwards20}. We believe our method of correcting for the dilution effect is well physically motivated based on our best knowledge of the companion star, consistent across all 3 instruments covering from 0.3 to 4.5 $\mu m$, and robust by integrating uncertainties on the parameters of the companion star. 

\begin{center}
\begin{deluxetable}{ccc}
\tablecaption{\textbf{\large{ATMO transit retrieval priors $\&$ Posteriors}}}
\tablehead{\colhead{Parameter} & \colhead{Priors} & \colhead{Posteriors}}
\startdata
\hline\hline 
log(Z/Z$_{\Sun}$)	&	$\mathcal{U}$(-2.8, 2.8)	&	$-2.309^{+0.574}_{-0.187}$	\\
R$_{pl}$(Jup)	&	$\mathcal{U}$(1.8565, 2.0519)	&	$1.945^{+0.004}_{-0.003}$	\\
log(K$_{IR}$)	&	$\mathcal{U}$(-5, -0.5)	&	$-2.198^{+0.004}_{-0.003}$	\\
log($\gamma$/IR)	&	$\mathcal{U}$(-4, 1.5)	&	$-1.558^{+0.750}_{-0.785}$	\\
beta	&	$\mathcal{U}$(0, 1.25)	&	$0.757^{+0.025}_{-0.026}$	\\
log(C/C$_{\Sun}$)	&	$\mathcal{U}$(-2.8, 2.8)	&	$-0.891^{+0.485}_{-0.656}$	\\
log(O/O$_{\Sun}$)	&	$\mathcal{U}$(-2.8, 2.8)	&	$-1.069^{+0.429}_{-0.481}$	\\
log(Na/Na$_{\Sun}$)	&	$\mathcal{U}$(-2.8, 2.8)	&	$0.649^{+0.511}_{-0.902}$	\\
log(Ti/Ti$_{\Sun}$)	&	$\mathcal{U}$(-2.8, 2.8)	&	$-0.365^{+0.819}_{-0.793}$	\\
log(V/V$_{\Sun}$)	&	$\mathcal{U}$(-2.8, 2.8)	&	$-0.713^{+0.510}_{-0.431}$	\\
log(Fe/Fe$_{\Sun}$)	&	$\mathcal{U}$(-2.8, 2.8)	&	$-0.275^{+0.866}_{-0.667}$	\\
\hline 
\enddata
\end{deluxetable}
\end{center}
\label{transit_priors}

\begin{center}
\begin{deluxetable}{ccc}
\tablecaption{\textbf{\large{ATMO eclipse retrieval priors $\&$ Posteriors}}}
\tablehead{\colhead{Parameter} & \colhead{Priors} & \colhead{Posteriors}}
\startdata
\hline\hline 
log(Z/Z$_{\Sun}$)	&	$\mathcal{U}$(-2.8, 2.8)	&	$-0.479^{+1.345}_{-1.169}$	\\
log(K$_{IR}$)	&	$\mathcal{U}$(-5, -0.5)	&	$-1.396^{+0.914}_{-1.138}$	\\
log($\gamma$/IR)	&	$\mathcal{U}$(-4, 1.5)	&	$0.459^{+0.522}_{-0.263}$	\\
beta	&	$\mathcal{U}$(0, 2)	&	$1.226^{+0.064}_{-0.061}$	\\
log(C/C$_{\Sun}$)	&	$\mathcal{U}$(-2.8, 2.8)	&	$0.658^{+0.885}_{-1.347}$	\\
log(O/O$_{\Sun}$)	&	$\mathcal{U}$(-2.8, 2.8)	&	$-0.567^{+1.646}_{-1.155}$	\\
\hline 
\enddata
\end{deluxetable}
\end{center}
\label{eclipse_priors}

\section{Analysis \& Interpretation}

After obtaining both the transit and eclipse spectra of WASP-76b, the next step is to physically interpret the spectra. Given different sets of parameters such as radius, metallicity, C/O ratio, temperature and aerosol properties, a model transit or eclipse spectrum can be generated via forward radiative transfer models based on transit and eclipse light path geometry. Running atmospheric models numerous times while varying the input parameters based on the goodness of fit of each combination and obtaining the posterior distribution of all parameters in the statistical framework is called a retrieval analysis \citep{irwin08, madhu14, line14, zhang18}. It allows us to obtain the best-fit physical parameters and their corresponding uncertainties. However, retrieval could be computational expensive depending on the complexity of individual forward model. Approximations such as a parametrized temperature-pressure (TP) profile, cloud scattering property or low-resolution opacity library are usually adopted to speed up the forward model and the retrieval. We performed retrieval analysis on WASP-76b using ATMO \citep{sing15} which is a MCMC algorithm based on forward radiative transfer models. In addition, we also ran a self-consistent PHOENIX \citep{lothringer18} model grid which uses radiative and chemical equilibrium. We used these two different models to cross validate and confirm the physical interpretation of the spectra. 

\begin{figure*}
  \includegraphics[width=1\textwidth,keepaspectratio]{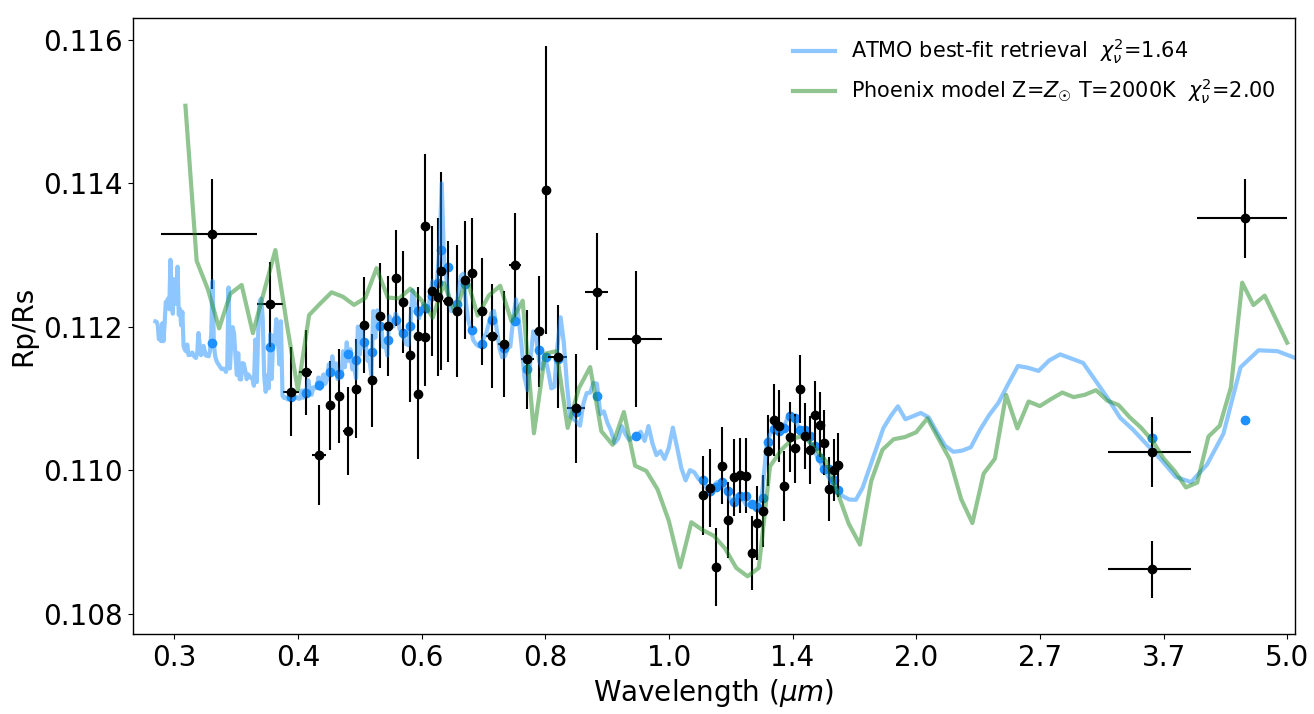}
  \caption{Transmission spectrum of WASP-76b overplotted two different best-fit models. The green line is the PHOENIX atmosphere model with equilibrium chemistry, solar metallicty and internal temperature of 200K. The blue line is the best-fit models from ATMO retrieval.}
  \label{fig:transit_phoenix}
\end{figure*}

\subsection{Strong metal absorbers in STIS G430L spectrum}

The WASP-76b spectrum shows a steep slope in the G430L spectrum. In other cooler hot Jupiters, the STIS blue part of the spectrum has been used to probe the Rayleigh scattering in the atmosphere as it usually exhibits larger transit depths and slopes down into longer wavelengths. However, 0.3 to 0.4 $\mu m$ of WASP-76b spectrum shows a much steeper slope compared to the rest of the spectrum which means one continuous Rayleigh scattering slope can not sufficiently explain the observed spectrum. To understand the origin of unexpected excess transit depth, we performed retrieval analysis with ATMO \citep{amundsen14, drummond16, goyal17, tremblin15, tremblin16}, which has been widely used before for retrieval analyses of transmission \citep{wakeford17} and emission \citep{evans17} spectra. We performed a cloud-free free-element equilibrium-chemistry retrieval with free abundance of specific species (C, O, Na, Ti, V, Fe) and a fitted TP profile (\ref{transit_priors}). All other elements were varied with a single metallicity parameter. ATMO is able to fit the STIS blue part of the spectrum with a solar Fe abundance and the best-fit model has a $\chi^2_{\nu}$ of 1.64. However, the first observed point extending from 0.29 to 0.37 $\mu m$ is still $\sim$2$\sigma$ higher than the ATMO model. 

The next modeling tool we applied is PHOENIX \citep{lothringer18} atmosphere forward model, It self-consistently solves layer by layer radiative transfer assuming chemical and radiative-convective equilibrium based on the irradiation received at the top of the atmosphere from the host star \citep{lothringer19}. PHOENIX is equipped with a comprehensive EUV-to-FIR opacity database of atomic opacity due to their importance in modeling stellar spectra, which makes it particularly suitable on predicting ultra-hot Jupiter atmospheres in the bluer wavelengths \citep{lothringer20}. We generated a grid of PHOENIX models with various metallicity, heat redistribution and internal temperature. The best-fit model (Fig. \ref{fig:transit_phoenix}) is at solar metallicity with a terminator temperature of 2000K which has a $\chi^2_{\nu}$ of 2. With the additional opacity from metals and molecules (Fe I, Fe II, Ti I, Ni I, Ca I, Ca II, and SiO) included in the PHOENIX model, it is able to fully fit the short-wavelength slope. However, it predicts larger absorption depth between the 0.4 to 0.5 $\mu m$ region which is likely due to the assumption of solar metallicity and elemental abundances. The lower than expected abundances of NUV absorbers such as TiO, V I and Fe I could be due to condensation and/or rain-out on the day-to-night terminator detected by \cite{ehrenreich20}. 

This similar feature of steep slope in the NUV has also been observed in WASP-121b \citep{evans17} with Sodium Hypochlorite (SH) proposed as the missing opacity source. With more recent observations \citep{sing19} with STIS E230M from 228 and 307 nm, multiple atomic lines including Mg II and Fe II have been detected and resolved in WASP-121b. This indicates neutral and ionized atomic metal lines are more likely to be the cause of the strong NUV absorption signatures in the STIS G430L spectrum. With both WASP-76b and WASP-121b showing strong NUV absorption features, neutral and ionized metals may exist in many more ultra-hot Jupiter atmospheres \citep{lothringer20}.

\begin{figure*}
  \includegraphics[width=\textwidth,keepaspectratio]{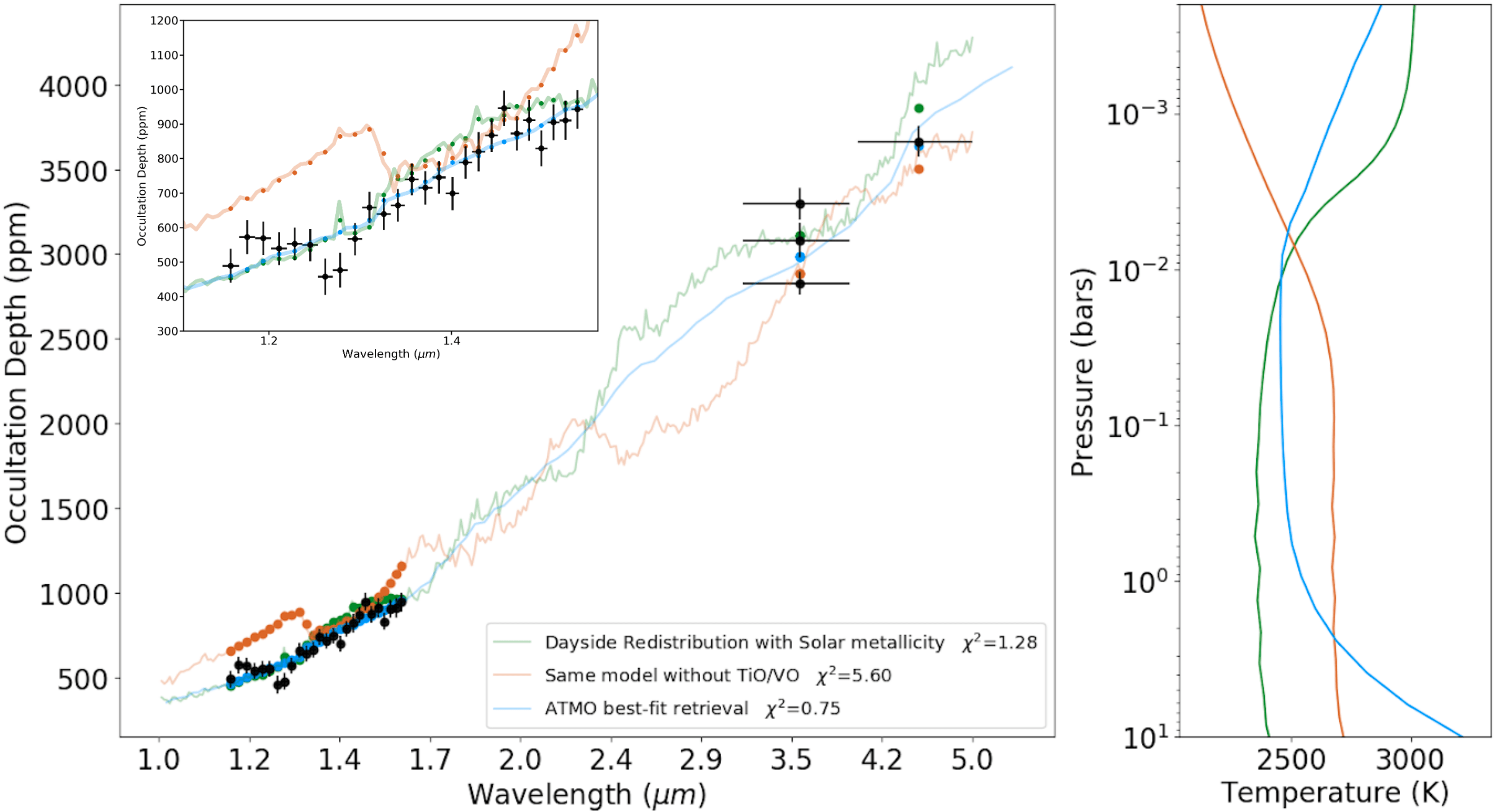}
  \caption{The eclipse spectrum (left panel) of WASP-76b overplotted with two PHOENIX (green and orange) models and one ATMO (blue) best-fit spectrum. The two PHOENIX models are both at solar metallicity with dayside heat redistribution, but one with TiO/VO and the other without. The comparison is to show the presence of TiO/VO is strongly favored by the data. The corresponding TP profiles are plotted in the right panel with matching colors to the three emission model spectra.}
  \label{fig:eclipse_phoenix}
\end{figure*}

\begin{figure}
  \includegraphics[width=0.5\textwidth,keepaspectratio]{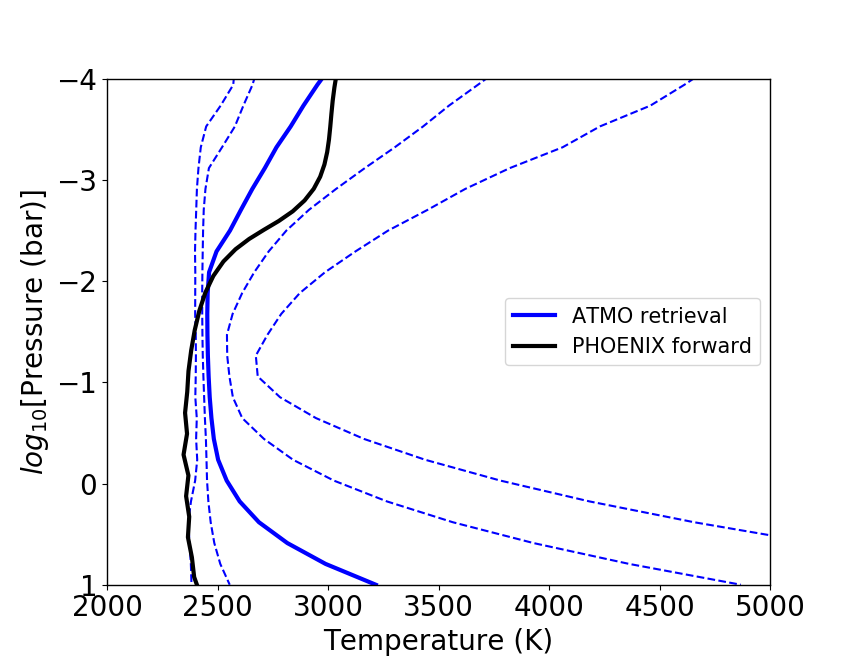}
  \caption{The TP profiles of ATMO retrieval (blue) and PHOENIX (black) atmosphere model for the emission spectrum. The dashed blue lines represent one and two sigma range for ATMO TP profiles.}
  \label{fig:TP_eclipse}
\end{figure}

\subsection{Detection of TiO and \texorpdfstring{H$_{2}$O}{H2O}}

\begin{figure*}
  \includegraphics[width=\textwidth,keepaspectratio]{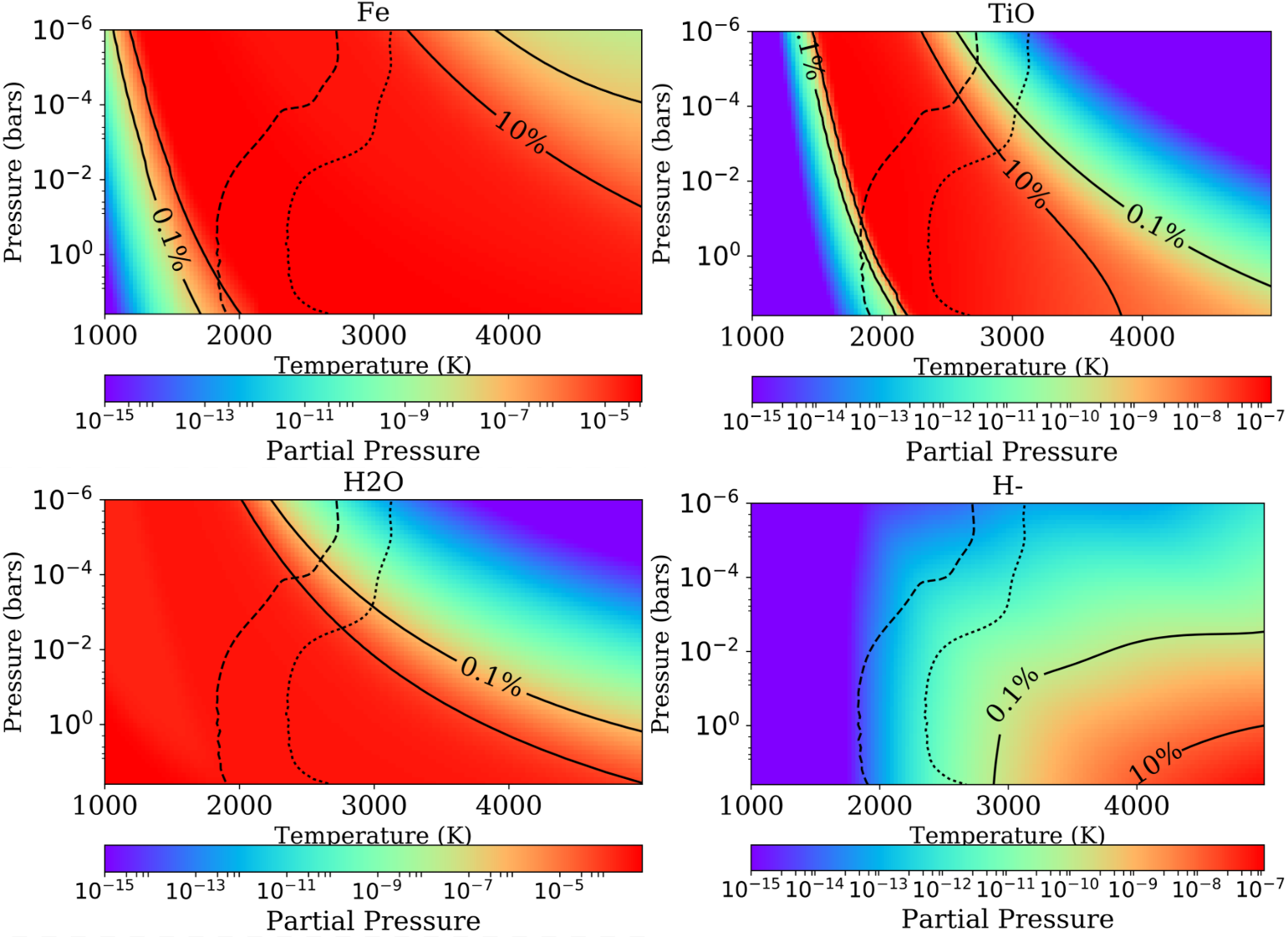}
  \caption{Partial pressure contours of four chemical species (Fe, TiO, H$_2$O and H-) overplotted with PHOENIX model TP profiles from transit (full heat redistribution, dashed lines) and eclipse (dayside heat redistribution, dotted lines) spectra. Gaseous Fe is abundant and presence in both the terminator and dayside regions of the planet across all pressure levels. TiO and water vapor exist in higher pressure regions but begin to dissociate in higher temperature and altitude layers. H- is limited in the cooler terminator but starts to show up on the hotter dayside.}
  \label{fig:partial_pressure}
\end{figure*}

We detected TiO and H$_{2}$O in the transmission spectrum of WASP-76b. The 0.4 to 1 $\mu m$ part of the spectrum where TiO opacity dominates shows significantly deeper transit depth ($\sim$500 ppm) compared to the WFC3/G141 spectrum. This feature is well explained by all two models with TiO absorption features. At this temperature range, TiO is expected to be in gaseous and abundant in the atmospheres as shown in the top-right panel of Fig. \ref{fig:partial_pressure}. Water vapor absorption feature at 1.4 $\mu m$ has also been observed in the spectrum, which is expected as thermal dissociation of water starts at temperatures $>$2500K (Bottom-left Fig. \ref{fig:partial_pressure}).

\subsection{Emission spectrum}

WASP-76b shows blackbody-like WFC3/G141 emission spectrum with muted water features but strong CO emission feature at Spitzer 4.5 $\mu m$ band. The best-fit PHOENIX model shows dayside heat-redistribution and solar metallicity assuming equilibrium chemistry. We also ran a comparison PHOENIX model with the same setup but excluding TiO/VO to demonstrate the data strongly favors the presence of gaseous TiO/VO, as the $\chi^2_{\nu}$ is larger by 4.32, which is consistent with our finding in the transmission spectrum. In addition, we performed ATMO free-element equilibrium chemistry retrieval similarly to the transmission spectrum, though isotopic scattering was also included along with the thermal emission. The resulting ATMO best-fit model is highly consistent compared to the PHOENIX model with both models showing similar emission spectra and TP profiles (See Fig. \ref{fig:TP_eclipse}). ATMO also favors solar metallicity in the retrieval posterior distribution (Fig. \ref{fig:arc_monitor_density_eclipse}) but with less certainty at the C/O ratio since the muted water feature limits the constrains on the oxygen abundance. Both models favor a dayside temperature range of 2500 to 2600K around 1 bar and an inverted TP profile with temperature increasing to around 3000K at 0.1 mbars. Water starts to dissociate at such high temperature and low pressure region of the atmosphere as shown in Figure \ref{fig:partial_pressure}, therefore we do not see prominent water emission features. At deeper levels ($\sim$1 bar) of the atmosphere, water vapor should still survive, but any absorption features will be obscured by the hotter continuum emission in the upper atmosphere layers. On the other hand, CO is able to survive in much higher altitude and temperature due to the strong triple bond structure. Indeed, we see clear CO emission features in the Spitzer 4.5 $\mu m$ band.

\subsection{Temperature inversion}

We found clear temperature inversion (Figure \ref{fig:TP_eclipse}) confirmed by ATMO and PHOENIX models. The Spitzer 4.5 $\mu m$ CO emission feature strongly favors an inverted TP profile with higher temperature CO gas presence in the upper atmospheres. The transmission spectrum also favors an inverted TP profile as the retrievals need the higher temperature at the low pressures to boost the scale heights and the size of spectral features to better match the data. Theories have indicated inversion is caused by a combination of optical absorbers such as TiO \citep{hubeny03, fortney08} and atomic metal absorption heating the upper layers with the lack of cooling from molecules like water \citep{lothringer18, gandhi19}. Our observed spectrum supports this paradigm with detection of TiO and atomic metal opacity in the transmission and muted water emission feature due to thermal dissociation at the highest altitudes. The detection of TiO and temperature inversion in the emission spectrum is also consistent with the independent analysis from \cite{edwards20} which reported similar findings.

\subsection{Model comparison}

ATMO best-fit model has the lower $\chi^2_{\nu}$ but PHOENIX generates a remarkable good fit in both transit and eclipse especially as only two parameters were varied in our grid of forward models. Retrieval frameworks find the best-fit spectrum through minimizing the likelihood which allows it to fine tune model parameters and better respond to smaller features in the data. Therefore, despite using an incomplete NUV opacity database, ATMO is able to produce better overall $\chi^2_{\nu}$ best-fit spectrum than the PHOENIX forward model. However, it is more important that ATMO and PHOENIX show good agreement on the general physical parameters including temperature structure, C/O ratio and chemical abundance. This gives us increased confidence in our conclusion, as both the retrieval and forward modeling methods agree.

\section{Summary and Conclusions}

We observed a combined total of 7 transits and 5 eclipses of the highly irradiated ultra-hot Jupiter WASP-76b using HST WFC3/STIS and $Spitzer$. After correcting for the dilution effect of a nearby companion star, and refitting the orbital parameters, we performed retrieval analysis on the transmission and emission spectra using ATMO and a PHOENIX grid. The results from these independent modeling tools are in generally good agreement with the biggest difference being the completeness of NUV opacity lines of each model. We demonstrated the importance of including all atomic and molecular metal lines in the NUV to fully explain the excess transit depth observed between 0.3 and 0.4 $\mu m$ in WASP-76b \citep{lothringer20}. Water vapor and TiO have also been directly detected in the transmission spectrum from STIS and WFC/G141 observations. Both transit and eclipse spectrum favor an inverted TP profile which is confirmed by both ATMO and PHOENIX models. The detection of TiO and ionized metals at the same time with an inverted TP profile are consistent with the theory of temperature inversion in ultra-hot Jupiters being caused by high altitude strong UV and optical absorbers heating up the upper layers. The lack of water emission due to dissociation at high temperature and altitude further drives the temperature inversion from the absence of cooling. 

This study of WASP-76b supports some of our current understanding of ultra-hot Jupiter such as their thermal structure while poses new questions about their heavy metals composition that have previously been mostly ignored. It is evident more NUV atmosphere observations of ultra-hot Jupiters are needed for a more complete understanding of these unique planets. HST is currently the only observatory capable of observing in the NUV wavelength which will not be accessible with JWST. WASP-76b along with other ultra-hot Jupiters will be great targets for future detailed NUV studies with HST.

\clearpage

\appendix

\providecommand{\bjdtdb}{\ensuremath{\rm {BJD_{TDB}}}}
\providecommand{\feh}{\ensuremath{\left[{\rm Fe}/{\rm H}\right]}}
\providecommand{\teff}{\ensuremath{T_{\rm eff}}}
\providecommand{\ecosw}{\ensuremath{e\cos{\omega_*}}}
\providecommand{\esinw}{\ensuremath{e\sin{\omega_*}}}
\providecommand{\msun}{\ensuremath{\,M_\Sun}}
\providecommand{\rsun}{\ensuremath{\,R_\Sun}}
\providecommand{\lsun}{\ensuremath{\,L_\Sun}}
\providecommand{\mj}{\ensuremath{\,M_{\rm J}}}
\providecommand{\rj}{\ensuremath{\,R_{\rm J}}}
\providecommand{\me}{\ensuremath{\,M_{\rm E}}}
\providecommand{\re}{\ensuremath{\,R_{\rm E}}}
\providecommand{\fave}{\langle F \rangle}
\providecommand{\fluxcgs}{10$^9$ erg s$^{-1}$ cm$^{-2}$}
\startlongtable
\begin{deluxetable*}{lccccccccc}
\tablecaption{Median values and 68$\%$ confidence interval for wasp-76b ExoFast v2 fit.}
\tablehead{\colhead{~~~Parameter} & \colhead{Units} & \multicolumn{8}{c}{Values}}
\startdata
\smallskip\\\multicolumn{2}{l}{Stellar Parameters:}&\smallskip\\
~~~~$M_*$\dotfill &Mass (\msun)\dotfill &$1.467^{+0.079}_{-0.081}$\\
~~~~$R_*$\dotfill &Radius (\rsun)\dotfill &$1.744^{+0.045}_{-0.042}$\\
~~~~$L_*$\dotfill &Luminosity (\lsun)\dotfill &$4.51^{+0.40}_{-0.37}$\\
~~~~$\rho_*$\dotfill &Density (cgs)\dotfill &$0.391^{+0.018}_{-0.024}$\\
~~~~$\log{g}$\dotfill &Surface gravity (cgs)\dotfill &$4.122^{+0.016}_{-0.020}$\\
~~~~$T_{\rm eff}$\dotfill &Effective Temperature (K)\dotfill &$6366^{+92}_{-90}$\\
~~~~$[{\rm Fe/H}]$\dotfill &Metallicity (dex)\dotfill &$0.204^{+0.093}_{-0.096}$\\
~~~~$[{\rm Fe/H}]_{0}$\dotfill &Initial Metallicity \dotfill &$0.315^{+0.086}_{-0.091}$\\
~~~~$Age$\dotfill &Age (Gyr)\dotfill &$1.86^{+0.75}_{-0.56}$\\
~~~~$EEP$\dotfill &Equal Evolutionary Phase \dotfill &$351.9^{+23}_{-7.5}$\\
\smallskip\\\multicolumn{2}{l}{Companion Star Parameters:}&$b$\smallskip\\
~~~~$P$\dotfill &Radius (\rsun)\dotfill &$0.795^{+0.055}_{-0.055}$\\
~~~~$T_{\rm eff}$\dotfill &Effective Temperature (K)\dotfill &$4850^{+150}_{-150}$\\
\smallskip\\\multicolumn{2}{l}{Planetary Parameters:}&$b$\smallskip\\
~~~~$P$\dotfill &Period (days)\dotfill &$1.80988158\pm0.00000030$\\
~~~~$R_P$\dotfill &Radius (\rj)\dotfill &$1.845^{+0.050}_{-0.046}$\\
~~~~$M_P$\dotfill &Mass (\mj)\dotfill &$0.910\pm0.042$\\
~~~~$T_C$\dotfill &Time of conjunction (\bjdtdb)\dotfill &$2456107.85494\pm0.00023$\\
~~~~$T_0$\dotfill &Optimal conjunction Time (\bjdtdb)\dotfill &$2457360.29300\pm0.00014$\\
~~~~$a$\dotfill &Semi-major axis (AU)\dotfill &$0.03304^{+0.00058}_{-0.00062}$\\
~~~~$i$\dotfill &Inclination (Degrees)\dotfill &$88.5^{+1.0}_{-1.2}$\\
~~~~$e$\dotfill &Eccentricity \dotfill &$0.016^{+0.018}_{-0.011}$\\
~~~~$\omega_*$\dotfill &Argument of Periastron (Degrees)\dotfill &$62^{+67}_{-82}$\\
~~~~$T_{eq}$\dotfill &Equilibrium temperature (K)\dotfill &$2231^{+37}_{-36}$\\
~~~~$\tau_{\rm circ}$\dotfill &Tidal circularization timescale (Gyr)\dotfill &$0.00260^{+0.00025}_{-0.00028}$\\
~~~~$K$\dotfill &RV semi-amplitude ($m/s$)\dotfill &$117.5\pm3.2$\\
~~~~$\log{K}$\dotfill &Log of RV semi-amplitude \dotfill &$2.070\pm0.012$\\
~~~~$R_P/R_*$\dotfill &Radius of planet in stellar radii \dotfill &$0.10873^{+0.00048}_{-0.00047}$\\
~~~~$a/R_*$\dotfill &Semi-major axis in stellar radii \dotfill &$4.078^{+0.060}_{-0.083}$\\
~~~~$\delta$\dotfill &Transit depth (fraction)\dotfill &$0.01182\pm0.00010$\\
~~~~$Depth$\dotfill &Flux decrement at mid transit \dotfill &$0.01182\pm0.00010$\\
~~~~$\tau$\dotfill &Ingress/egress transit duration (days)\dotfill &$0.01576^{+0.00043}_{-0.00018}$\\
~~~~$T_{14}$\dotfill &Total transit duration (days)\dotfill &$0.15636^{+0.00054}_{-0.00048}$\\
~~~~$T_{FWHM}$\dotfill &FWHM transit duration (days)\dotfill &$0.14051^{+0.00039}_{-0.00038}$\\
~~~~$b$\dotfill &Transit Impact parameter \dotfill &$0.103^{+0.084}_{-0.071}$\\
~~~~$b_S$\dotfill &Eclipse impact parameter \dotfill &$0.105^{+0.085}_{-0.072}$\\
~~~~$\tau_S$\dotfill &Ingress/egress eclipse duration (days)\dotfill &$0.01611^{+0.00075}_{-0.00051}$\\
~~~~$T_{S,14}$\dotfill &Total eclipse duration (days)\dotfill &$0.1584^{+0.0065}_{-0.0033}$\\
~~~~$T_{S,FWHM}$\dotfill &FWHM eclipse duration (days)\dotfill &$0.1423^{+0.0059}_{-0.0029}$\\
~~~~$\delta_{S,3.6\mu m}$\dotfill &Blackbody eclipse depth at 3.6$\mu$m (ppm)\dotfill &$2037^{+56}_{-47}$\\
~~~~$\delta_{S,4.5\mu m}$\dotfill &Blackbody eclipse depth at 4.5$\mu$m (ppm)\dotfill &$2407^{+57}_{-48}$\\
~~~~$\rho_P$\dotfill &Density (cgs)\dotfill &$0.180^{+0.012}_{-0.013}$\\
~~~~$logg_P$\dotfill &Surface gravity \dotfill &$2.822^{+0.020}_{-0.023}$\\
~~~~$\Theta$\dotfill &Safronov Number \dotfill &$0.02219^{+0.00088}_{-0.00087}$\\
~~~~$\fave$\dotfill &Incident Flux (\fluxcgs)\dotfill &$5.62^{+0.38}_{-0.35}$\\
~~~~$T_P$\dotfill &Time of Periastron (\bjdtdb)\dotfill &$2456107.73^{+0.34}_{-0.40}$\\
~~~~$T_S$\dotfill &Time of eclipse (\bjdtdb)\dotfill &$2456108.7641^{+0.015}_{-0.0099}$\\
~~~~$T_A$\dotfill &Time of Ascending Node (\bjdtdb)\dotfill &$2456109.2186^{+0.015}_{-0.0084}$\\
~~~~$T_D$\dotfill &Time of Descending Node (\bjdtdb)\dotfill &$2456108.3060^{+0.0077}_{-0.012}$\\
~~~~$e\cos{\omega_*}$\dotfill & \dotfill &$0.0036^{+0.013}_{-0.0086}$\\
~~~~$e\sin{\omega_*}$\dotfill & \dotfill &$0.006^{+0.021}_{-0.010}$\\
~~~~$M_P\sin i$\dotfill &Minimum mass (\mj)\dotfill &$0.910\pm0.042$\\
~~~~$M_P/M_*$\dotfill &Mass ratio \dotfill &$0.000593^{+0.000020}_{-0.000019}$\\
~~~~$d/R_*$\dotfill &Separation at mid transit \dotfill &$4.050^{+0.098}_{-0.16}$\\
~~~~$P_T$\dotfill &A priori non-grazing transit prob \dotfill &$0.2201^{+0.0090}_{-0.0052}$\\
~~~~$P_{T,G}$\dotfill &A priori transit prob \dotfill &$0.2738^{+0.011}_{-0.0065}$\\
~~~~$P_S$\dotfill &A priori non-grazing eclipse prob \dotfill &$0.2165^{+0.0024}_{-0.0012}$\\
~~~~$P_{S,G}$\dotfill &A priori eclipse prob \dotfill &$0.2692^{+0.0032}_{-0.0015}$\\
\enddata
\label{exofast}
\end{deluxetable*}

\startlongtable
\begin{center}
\begin{deluxetable*}{ccccc}
\tablecaption{\textbf{\large{WASP-76b transit spectrum}}}
\tablehead{\colhead{Wavelength midpoint ($\mu$m)} & \colhead{Bin width ($\mu$m)} & \colhead{Rp/Rs} & \colhead{Rp/Rs uncertainty} & \colhead{Dilution factor}}
\startdata
\hline\hline 
0.33000	&	0.04000	&	0.11329	&	0.00074	&	1.0086	\\
0.38250	&	0.01250	&	0.11231	&	0.00058	&	1.00807	\\
0.40315	&	0.00815	&	0.11110	&	0.00063	&	1.01406	\\
0.41815	&	0.00685	&	0.11137	&	0.00061	&	1.01325	\\
0.43250	&	0.00750	&	0.11021	&	0.00070	&	1.01498	\\
0.44500	&	0.00500	&	0.11090	&	0.00062	&	1.01766	\\
0.45500	&	0.00500	&	0.11104	&	0.00064	&	1.02015	\\
0.46500	&	0.00500	&	0.11055	&	0.00060	&	1.02093	\\
0.47500	&	0.00500	&	0.11114	&	0.00071	&	1.02133	\\
0.48500	&	0.00500	&	0.11203	&	0.00069	&	1.0231	\\
0.49500	&	0.00500	&	0.11126	&	0.00066	&	1.02192	\\
0.50500	&	0.00500	&	0.11215	&	0.00074	&	1.02105	\\
0.51500	&	0.00500	&	0.11202	&	0.00070	&	1.01997	\\
0.52500	&	0.00500	&	0.11268	&	0.00067	&	1.02327	\\
0.53500	&	0.00500	&	0.11235	&	0.00070	&	1.02497	\\
0.54500	&	0.00500	&	0.11161	&	0.00068	&	1.02527	\\
0.55500	&	0.00500	&	0.11187	&	0.00066	&	1.02664	\\
0.56500	&	0.00500	&	0.11186	&	0.00069	&	1.02765	\\
0.55500	&	0.00500	&	0.11107	&	0.00089	&	1.02664	\\
0.56500	&	0.00500	&	0.11341	&	0.00099	&	1.02765	\\
0.57500	&	0.00500	&	0.11250	&	0.00088	&	1.02814	\\
0.58390	&	0.00390	&	0.11241	&	0.00108	&	1.02894	\\
0.58955	&	0.00175	&	0.11278	&	0.00139	&	1.0275	\\
0.59915	&	0.00785	&	0.11236	&	0.00082	&	1.02961	\\
0.61350	&	0.00650	&	0.11223	&	0.00091	&	1.03001	\\
0.62500	&	0.00500	&	0.11266	&	0.00080	&	1.03027	\\
0.63750	&	0.00750	&	0.11275	&	0.00077	&	1.03116	\\
0.65250	&	0.00750	&	0.11221	&	0.00076	&	1.03279	\\
0.67000	&	0.01000	&	0.11187	&	0.00071	&	1.03279	\\
0.69000	&	0.01000	&	0.11176	&	0.00074	&	1.03355	\\
0.71000	&	0.01000	&	0.11286	&	0.00072	&	1.03411	\\
0.73250	&	0.01250	&	0.11155	&	0.00070	&	1.03511	\\
0.75475	&	0.00975	&	0.11194	&	0.00075	&	1.03628	\\
0.76825	&	0.00375	&	0.11391	&	0.00197	&	1.03683	\\
0.79100	&	0.01900	&	0.11158	&	0.00072	&	1.03747	\\
0.82925	&	0.01925	&	0.11087	&	0.00079	&	1.03862	\\
0.87350	&	0.02500	&	0.11249	&	0.00082	&	1.04058	\\
0.96425	&	0.06575	&	0.11183	&	0.00094	&	1.04236	\\
1.14250	&	0.00930	&	0.10965	&	0.00055	&	1.04738	\\
1.16110	&	0.00930	&	0.10976	&	0.00055	&	1.04821	\\
1.17970	&	0.00930	&	0.10865	&	0.00054	&	1.04882	\\
1.19830	&	0.00930	&	0.11007	&	0.00054	&	1.04922	\\
1.21690	&	0.00930	&	0.10930	&	0.00053	&	1.04999	\\
1.23550	&	0.00930	&	0.10990	&	0.00054	&	1.05059	\\
1.25410	&	0.00930	&	0.10993	&	0.00052	&	1.05126	\\
1.27270	&	0.00930	&	0.10992	&	0.00052	&	1.05254	\\
1.29130	&	0.00930	&	0.10884	&	0.00051	&	1.05325	\\
1.30990	&	0.00930	&	0.10926	&	0.00051	&	1.05296	\\
1.32850	&	0.00930	&	0.10943	&	0.00050	&	1.05367	\\
1.34710	&	0.00930	&	0.11027	&	0.00050	&	1.05453	\\
1.36570	&	0.00930	&	0.11071	&	0.00050	&	1.05525	\\
1.38430	&	0.00930	&	0.11062	&	0.00050	&	1.056	\\
1.40290	&	0.00930	&	0.10978	&	0.00049	&	1.05647	\\
1.42150	&	0.00930	&	0.11047	&	0.00049	&	1.05618	\\
1.44010	&	0.00930	&	0.11030	&	0.00048	&	1.05748	\\
1.45870	&	0.00930	&	0.11113	&	0.00047	&	1.05799	\\
1.47730	&	0.00930	&	0.11048	&	0.00046	&	1.05901	\\
1.49590	&	0.00930	&	0.11029	&	0.00047	&	1.05928	\\
1.51450	&	0.00930	&	0.11078	&	0.00047	&	1.06108	\\
1.53310	&	0.00930	&	0.11064	&	0.00045	&	1.06235	\\
1.55170	&	0.00930	&	0.11039	&	0.00045	&	1.06346	\\
1.57030	&	0.00930	&	0.10974	&	0.00044	&	1.06339	\\
1.58890	&	0.00930	&	0.11000	&	0.00043	&	1.06444	\\
1.60750	&	0.00930	&	0.11008	&	0.00044	&	1.06581	\\
3.55000	&	0.37500	&	0.11026	&	0.00048	&	1.06895	\\
3.55000	&	0.37500	&	0.10862	&	0.00041	&	1.06895	\\
4.49300	&	0.50750	&	0.11351	&	0.00056	&	1.06709	\\
\hline 
\enddata
\end{deluxetable*}
\end{center}
\label{transit_spectrum}

\startlongtable
\begin{center}
\begin{deluxetable*}{ccccc}
\tablecaption{\textbf{\large{WASP-76b emission spectrum}}}
\tablehead{\colhead{Wavelength midpoint ($\mu$m)} & \colhead{Bin width ($\mu$m)} & \colhead{Occultation Depth (ppm)} & \colhead{Uncertainty (ppm)} & \colhead{Dilution factor}}
\startdata
\hline\hline 
1.1518	&	0.0093	&	490	&	49	&	1.04775	\\
1.1704	&	0.0093	&	573	&	49	&	1.04869	\\
1.1890	&	0.0093	&	570	&	48	&	1.04894	\\
1.2076	&	0.0093	&	540	&	48	&	1.04953	\\
1.2262	&	0.0093	&	553	&	47	&	1.05039	\\
1.2448	&	0.0093	&	550	&	47	&	1.05078	\\
1.2634	&	0.0093	&	458	&	53	&	1.05168	\\
1.2820	&	0.0093	&	477	&	50	&	1.05360	\\
1.3006	&	0.0093	&	567	&	47	&	1.05280	\\
1.3192	&	0.0093	&	659	&	45	&	1.05328	\\
1.3378	&	0.0093	&	640	&	47	&	1.05402	\\
1.3564	&	0.0093	&	665	&	47	&	1.05485	\\
1.3750	&	0.0093	&	740	&	46	&	1.05578	\\
1.3936	&	0.0093	&	716	&	48	&	1.05623	\\
1.4122	&	0.0093	&	746	&	47	&	1.05631	\\
1.4308	&	0.0093	&	699	&	49	&	1.05671	\\
1.4494	&	0.0093	&	789	&	47	&	1.05781	\\
1.4680	&	0.0093	&	820	&	57	&	1.05830	\\
1.4866	&	0.0093	&	868	&	48	&	1.05939	\\
1.5052	&	0.0093	&	947	&	51	&	1.06001	\\
1.5238	&	0.0093	&	874	&	50	&	1.06173	\\
1.5424	&	0.0093	&	912	&	59	&	1.06301	\\
1.5610	&	0.0093	&	830	&	52	&	1.06356	\\
1.5796	&	0.0093	&	906	&	52	&	1.06397	\\
1.5982	&	0.0093	&	911	&	57	&	1.06442	\\
1.6168	&	0.0093	&	943	&	56	&	1.06572	\\
3.5500	&	0.3750	&	2827	&	69	&	1.06895	\\
3.5500	&	0.3750	&	3082	&	102	&	1.06895	\\
3.5500	&	0.3750	&	3299	&	94	&	1.06895	\\
4.4930	&	0.5075	&	3665	&	89	&	1.06709	\\
\hline 
\enddata
\end{deluxetable*}
\end{center}
\label{emission_spectrum}

\begin{figure}
\centering
  \makebox[\textwidth][c]{\includegraphics[width=0.8\textwidth]{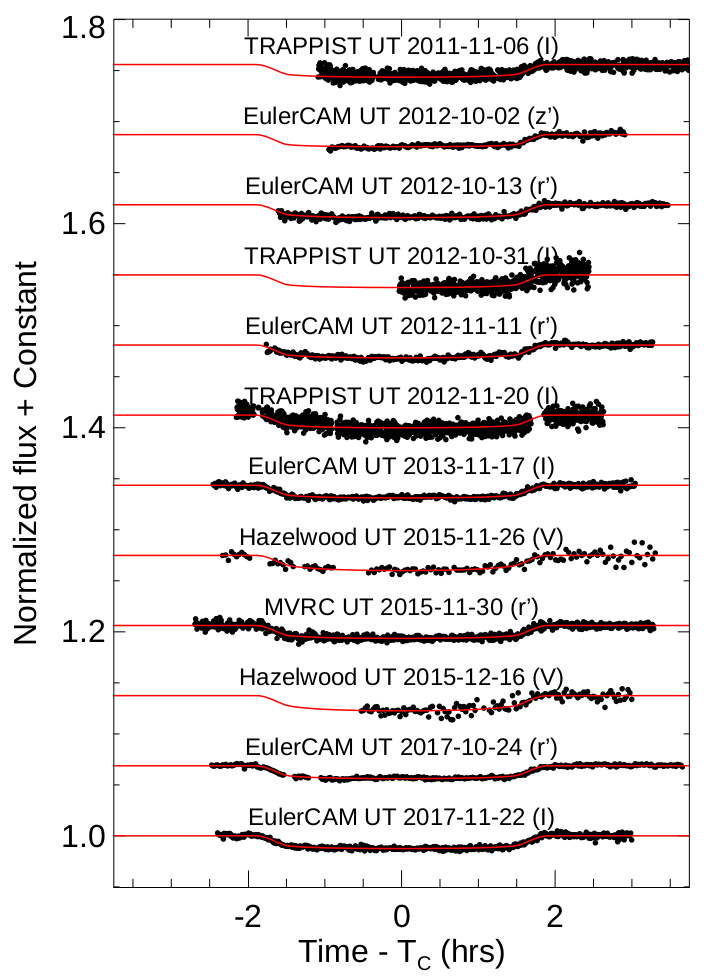}}
  \caption{Transit lightcurves used in EXOFASTv2 fit.}
  \label{fig:exofast_transits}
\end{figure}

\begin{figure*}
  \includegraphics[width=\textwidth, keepaspectratio]{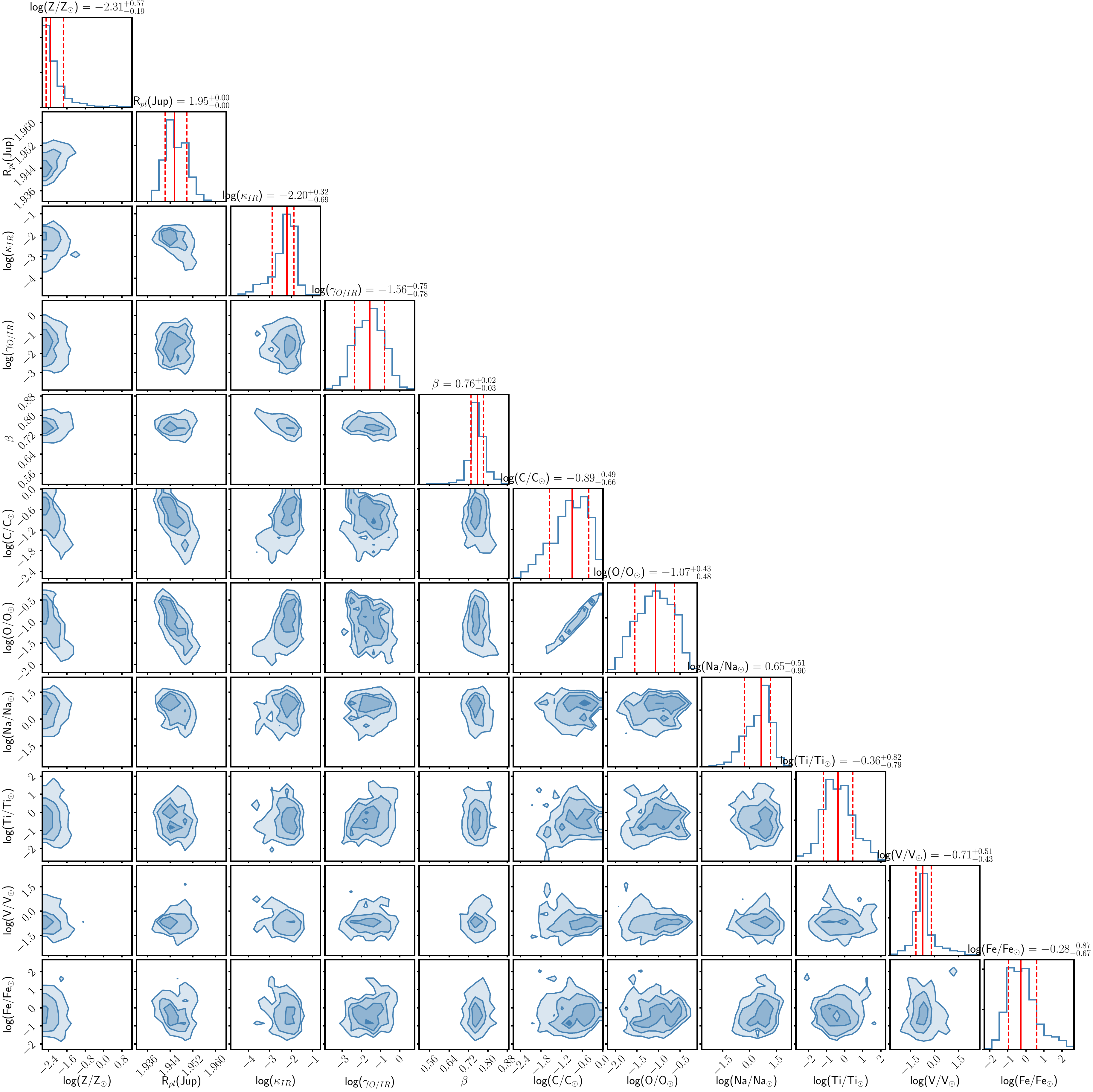}
  \caption{The posterior distribution of ATMO retrieval of the transmission spectrum. Six chemical elements (C, O, Na, Ti, V, Fe) are allowed to vary freely with everything else scale with solar metallicity. All retrieved elemental abundance are consistent with solar value to one sigma.}
  \label{fig:arc_monitor_density}
\end{figure*}

\begin{figure*}
  \includegraphics[scale=0.8, angle=-90]{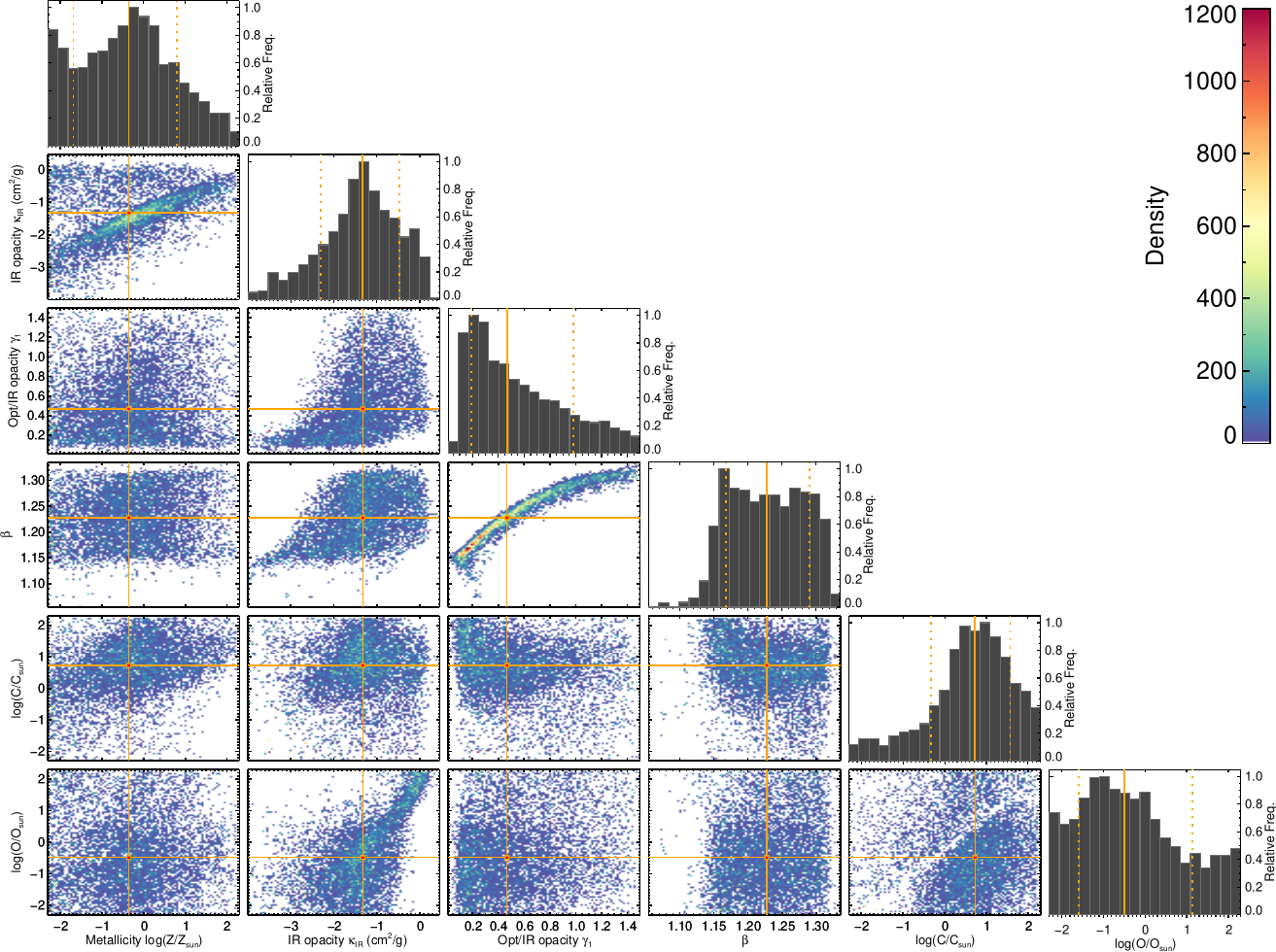}
  \caption{The posterior distribution of ATMO retrieval of the emission spectrum. The carbon and oxygen abundances are poorly constrained due to a muted water feature in the WFC3/G141 band. Retrieved solar metallicity is consistent with results from the transmission spectrum and the PHOENIX models.}
  \label{fig:arc_monitor_density_eclipse}
\end{figure*}
\end{document}